\documentclass[preprint, preprintnumbers,showpacs, showkeys, aps, prd, amsmath, amsfonts, amssymb, eqsecnum, nofootinbib, floatfix, secnumarabic]{revtex4-1}
\usepackage{amsmath,amssymb,graphicx}
\usepackage{slashed}
\usepackage{epsf}
\usepackage{float}
\pdfoutput=1

\newcommand{\nc}{\newcommand}
\nc{\postscript}[2]
{\setlength{\epsfxsize}{#2\hsize}\centerline{\epsfbox{#1}}}
\nc{\non}{\nonumber}
\nc{\hc}{\hbox {h.c.}} \nc{\re}{\hbox {Re}} 
\nc{\mev}{\hbox {MeV}} \nc{\gev}{\;\hbox {GeV}} \nc{\tev}{\;\hbox {TeV}}
\def\lsim{\mathrel{\raise.3ex\hbox{$<$\kern-.75em\lower1ex\hbox{$\sim$}}}}
\def\gsim{\mathrel{\raise.3ex\hbox{$>$\kern-.75em\lower1ex\hbox{$\sim$}}}}

\nc{\etal}{{\it et al.}}
\nc{\Lsp}{\;\;\;\;\;\;\;\;\;\;}  \nc{\LLLsp}{\lspace \lspace}
\nc{\lsp}{\;\;\;\;\;\;}
\nc{\spac}{\;\;\;}
\nc{\noi}{\noindent}
\nc{\beq}{\begin{equation}}   \nc{\eeq}{\end{equation}}
\nc{\bea}{\begin{eqnarray}}   \nc{\eea}{\end{eqnarray}}
\nc{\baa}{\begin{array}}      \nc{\eaa}{\end{array}}
\nc{\bit}{\begin{itemize}}    \nc{\eit}{\end{itemize}}
\nc{\ben}{\begin{enumerate}}  \nc{\een}{\end{enumerate}}
\nc{\bce}{\begin{center}}     \nc{\ece}{\end{center}}

\def\ie{{\it i.e.,}}

\def\cf{{\it cf.}}

\def\eps{\epsilon}

\def\sq2{\sqrt{2}}

\def\ph{\varphi}

\def\m4{m^4(\ph)}
\def\mn2{m_n^2}

\def\v5{V^{(5)}}

\def\baa{\begin{array}}
\def\eaa{\end{array}}

\begin{document}

\begin{flushright}
 \mbox{\normalsize \rm CUMQ/HEP 181}\\
\end{flushright}

\vskip -10pt

\title{Fermion Masses and Mixing in General Warped Extra Dimensional Models}

\author{Mariana Frank$^1$\footnote{mariana.frank@concordia.ca}}
\author{Cherif Hamzaoui$^2$\footnote{hamzaoui.cherif@uqam.ca}}
\author{Nima Pourtolami$^1$\footnote{n\_pour@live.concordia.ca}}
\author{Manuel Toharia$^1$\footnote{mtoharia@physics.concordia.ca}}
\affiliation{$^1 $ Department of Physics, Concordia University 
7141 Sherbrooke St. West, Montreal, Quebec,   CANADA H4B 1R6}
\affiliation{$^2$ Groupe de Physique Th\'eorique des Particules,
D\'epartement des Sciences de la Terre et de L'Atmosph\`ere,
Universit\'e du Qu\'ebec \`a Montr\'eal, Case Postale 8888, Succ. Centre-Ville,
 Montr\'eal, Qu\'ebec, Canada, H3C 3P8}


\begin{abstract}

We analyze fermion masses and mixing in a general warped extra
dimensional model, where all the Standard Model (SM) fields, including the Higgs, are
allowed to propagate in the bulk.
In this context, a slightly broken flavor symmetry imposed universally
on all fermion fields, without distinction, can generate   
the full flavor structure of the SM, including quarks, charged leptons
and neutrinos.
For quarks and charged leptons, the exponential
sensitivity of their wave-functions to small flavor breaking
effects yield naturally hierarchical masses and mixing 
as it is usual  in warped models with fermions in the bulk.
In the  neutrino sector, the exponential wave-function factors can be
flavor-blind and thus insensitive to the small flavor symmetry breaking effects,
directly linking their masses and mixing angles to
the flavor symmetric structure of the 5D neutrino Yukawa couplings.
The Higgs must be localized in the bulk and the model 
is naturally more successful in generalized warped scenarios where the
metric background solution is different than $AdS_5$.
We study these features in two simple frameworks, flavor
complimentarily, and flavor democracy, which provide 
specific predictions and correlations between quarks and leptons,
testable as more precise data in the neutrino sector becomes 
available. 

\end{abstract}


\maketitle



\section{Introduction}
\label{sec:intro}
The discovery  of a SM-like Higgs boson at the LHC with a mass of 125 GeV was a
huge step forward in confirming the validity of the Standard Model
(SM) and probing the electroweak symmetry breaking mechanism.
But despite its experimental success, the SM still fails to provide an
explanation, among other things, for the origin and observed pattern of fermion masses and
mixings.

In the quark sector, the masses  are extremely hierarchical,
with the top much heavier than the rest of the quarks and with a
strong ordering. In the up sector, the masses are separated by three
orders of magnitude, 
while in the down sector the mass ratios are separated by two orders
of magnitude.
The quarks also exhibit mixing patterns 
given by three small mixing angles and a large (CP) phase. In the lepton
sector, the charged lepton masses obey a similar 
hierarchical pattern as the down-type quarks.
On the other hand, though not known exactly, neutrino masses are known to be very small 
and their square mass differences imply a closer mass pattern, $\Delta
m_{31(32)}^2 : \Delta m_{21}^2\sim 10^2:1$.
Neutrinos appear to mix maximally and this has
been long-seen as pointing towards a different flavor origin between quarks
and leptons, and also to the necessity of introducing new
physics. 
After the successful measurement of the neutrino mixing angle
$\theta_{13}$ by the Daya Bay \cite{An:2012eh,An:2013uza} , 
T2K \cite{Abe:2011sj,Abe:2014ugx}, MINOS
\cite{Adamson:2011ig,Adamson:2014vgd}, RENO \cite{Ahn:2012nd} 
and Double Chooz \cite{Abe:2012tg,Abe:2014bwa} Collaborations, the
determination of the neutrino mass hierarchy has become a priority for
theoretical studies and for future neutrino experiments.  A great deal of
theoretical work in this area has been trying to provide answers,
based on such diverse frameworks as see-saw mechanisms 
\cite{Minkowski:1977sc,Supergravity:GellMann80,Yanagida79,Glashow80,Mohapatra:1979ia,Schechter:1980gr,Lazarides:1980nt},
Abelian \cite{Frampton:2002yf,Lavoura:2004tu,Dev:2010if} and non-Abelian 
\cite{Lam:2008rs,Lam:2008sh,Grimus:2009pg,Ge:2011ih,He:2011kn,Toorop:2011jn,Ge:2011qn,
deAdelhartToorop:2011re,He:2012yt,Hernandez:2012ra,Lam:2012ga,Hernandez:2012sk,
Holthausen:2012wt,Hu:2012ei,Hernandez:2013vya,King:2013vna,Hagedorn:2013nra,Lavoura:2014kwa} symmetries
imposed on the leptonic sector (both charged and neutral), and many
texture structures for leptonic mass matrices, including modifications
of accepted paradigms, such as tri-bimaximal 
\cite{Harrison:1999cf,Harrison:2002er,Xing:2002sw,Harrison:2002kp,Harrison:2003aw,He:2003rm}, 
bi-maximal
\cite{Vissani:1997pa,Barger:1998ta,Baltz:1998ey,Stancu:1999ct,Georgi:1998bf,Li:2004nh} 
and democratic 
\cite{Fritzsch:1995dj,Fritzsch:1998xs,Fritzsch:1999im} 
neutrino mixing matrices. While
various  attempts to unify the description of quarks and leptons
already exist (mostly based on quark-lepton complementarity
\cite{Zhang:2012pv,Zhang:2012zh,Li:2011ag,Kang:2011tv,Ahn:2011yj,Shimizu:2010pg,Patel:2010hr,
Barranco:2010we,Zheng:2010kp,Xing:2009eg,Frampton:2008ep,Rodejohann:2008zz,Plentinger:2007px,
Picariello:2007ss}), 
an attractive possibility would be that quarks and
leptons obey the same symmetry at a higher scale, which is then
slightly broken at lower scales, yielding different patterns for masses and mixing for
the quarks/leptons than for the neutrinos.  This is the scenario we
plan to investigate here, in the context of warped extra dimensions, where 
small symmetry breaking terms have very different effects on quarks and leptons due to the geometry
of the model. 

Introducing a warped extra dimension provides an elegant way to
address both the hierarchy problem (to stabilize the Higgs mass
against large radiative corrections) and the fermion flavor hierarchy problem. 
These models were first proposed to deal with the
hierarchy problem of the  SM \cite{Randall:1999ee,Randall:1999vf}, where  
introducing an extra
dimension produced a five dimensional anti-de Sitter  ($AdS_5$)
geometry bounded by two hard walls (branes)  along the  extra
dimension, referred to as the Planck (or UV) brane and the  TeV (or
IR) brane.  Allowing for exponential modulation from the  gravity scale down to the  weak scale  along this compact
extra dimension 
\cite{Randall:1999ee,Randall:1999vf}, naturally yields the   
weak-Planck mass
hierarchy. 

The original Randall-Sundrum (RS) model localized all SM fields on the IR brane, leading
to severe flavor violation bounds on the new physics scale. It was later shown that if the  fermions were
allowed to propagate in the bulk of the extra dimension
\cite{Grossman:1999ra,Gherghetta:2000qt,Davoudiasl:1999tf,Chang:1999nh,Casagrande:2010si,Huber:2003tu}, the same 
model could  address the flavor hierarchy problem of the SM as
well. The model thus emerged as a geometric theory of
flavor. By localizing the Higgs on the IR brane with anarchic order-one
couplings to the bulk fermions, the profiles of the fermion zero-modes
can be adjusted to reproduce the observed Yukawa couplings in the low
energy theory. Since the first  and second generation fermions are
localized towards the UV brane, they inherit substantial flavor
protection from the RS-GIM mechanism \cite{Agashe:2004cp}. However,
by allowing the SM fields to propagate in the bulk, from an
effective 4D point of view, a tower of KK fermions exists for all the
SM fields, yielding enhanced contributions to  electroweak and flavor
observables in the SM 
\cite{Agashe:2003zs,Burdman:2002gr,Burdman:2003nt,Huber:2003tu,
Carena:2004zn,Carena:2003fx,Delgado:2007ne,Agashe:2004cp,Quiros:2013yaa}. This
effect imposes  a stringent bound on the scale of new physics of some
$\sim 10$ TeV \cite{Csaki:2008zd,Agashe:2008uz} and hence  renders these models
completely out of the reach of current experiments.  Some solutions
were proposed to relieve these restrictions. One way was to
extend the gauge symmetry of the model by introducing an $SU(2)_R$
gauge sector with custodial protection \cite{Agashe:2006at,
  Huber:2003tu, Agashe:2003zs}. Here the basic idea is to align the
down-type Yukawa couplings using the additional symmetry, so that the
primary sources of intergenerational mixing are the up-type Yukawa
couplings. Since the dominant constraints on FCNCs come from the
down-type sector, the constraint on KK masses is substantially
relaxed. A different approach to address the issue is through a slight
modification  of the warping factor along the extra dimension, allowing
it to deviate slightly from the $AdS_5$  metric 
\cite{Falkowski:2008fz,Batell:2008me,MertAybat:2009mk,Cabrer:2009we,Cabrer:2010si,Cabrer:2011fb}. This deviation is
such that the warping  is more drastic near the TeV brane, while the
background becomes more $AdS_5$-like near the  Planck brane. These
type of metric solutions can help suppress dangerous contributions to
the electroweak and flavor observables by reducing the constraints on
new physics down to about $\sim 1$ TeV.

Warped extra dimensional models were shown to provide new contributions 
 to the Higgs production rate through gluon fusion which could conflict with the collider data \cite{Casagrande:2010si,Azatov:2010pf,Frank:2013un,Frank:2013qma}.

Interestingly, in the modified 5D metric scenarios the region of
parameter space in which the dangerous contributions to flavor
and electroweak precision observables are small is the same as the region where 
 the new contributions to the Higgs production cross section are
also small (and thus safe) \cite{Frank:2013qma}. Moreover, this is achievable only when the
Higgs field in these models is allowed to leak considerably into the bulk.

In the context of warped extra dimensional models with ${\cal O}(1)$
5D Yukawa couplings and with no {\it a priori} structure  (\ie ~flavor
anarchy), one can easily generate the hierarchical
structure of the quark and charged lepton sectors \cite{Csaki:2008qq}
while, due to large mixing angles, the neutrino sector must be treated
differently. In particular, in \cite{Agashe:2008fe}, it was shown that
if the Higgs field leaks sufficiently into the bulk it is possible that the
(exponentially small) neutrino wave functions become independent on
the flavor structure of the 5D neutrino mass parameters ($c^i_\nu$), 
and thus the 4D neutrino flavor structure depends directly on the
flavor structure of the 5D neutrino Yukawa couplings.

In a previous work \cite{Frank:2014aca}, we proposed  a unified picture of fermion
masses and mixings in the context of a warped extra dimensional
model with $AdS_5$ background metric, and with all the SM fields in the bulk, including the Higgs.
In that picture, the  same  flavor symmetric structure is imposed on
all the fermions of the SM, including neutrinos.
Small flavor breaking effects are exponentially enhanced in the quark
and charged lepton sectors, thus producing hierarchical masses and
mixings. With a sufficiently delocalized Higgs field, the neutrino wave functions
are flavor-blind and the flavor structure is governed by the 5D
neutrino Yukawa flavor structure.

In this work, we revisit this idea in the context of the
modified $AdS_5$ metric solutions. We show that the SM masses and
mixing can be generated successfully, and the mass generation in the neutrino sector appears
to be more natural than in the case where a pure $AdS_5$
background metric is assumed.

Our work is organized as follows. We summarize the features of the modified $AdS_5$ 
model in Section \ref{sec:01}, with particular emphasis on fermion mass
generation. We explore an explicit implementation of the scenario, flavor complementarity, in
Section \ref{sec:02} and another, of a democratic flavor
symmetry, in Section \ref{sec:03}. We summarize our results and conclude in Section
\ref{sec:summary}. We leave the details  
 of some calculations for
the Appendix \ref{app4}.


\section{Fermion masses in warped space}
\label{sec:01}
We consider a 5D warped space with the extra dimension compactified and allow all SM fields to propagate in the following generalized warped space-time metric:
\beq
ds^2 = e^{-2 A(y)} \eta_{\mu\nu}dx^{\mu}dx^{\nu} - dy^2,
\eeq
$\eta_{\mu\nu}={\rm diag}(-1,1, 1, 1)$ being the flat metric. The 5-th dimension, $y$, is bounded by two branes localized at
$y=0$ and $y=y_1$ and $A(y)$ is a model-dependent function. As
mentioned in the Introduction, some generalized warped models can be
safe from precision electroweak tests and flavor bounds for very low KK masses. 
Motivated by this, we consider a modified $AdS_5$ scenario with the
following warp exponent \cite{Cabrer:2010si,Carmona:2011rd}: 
\beq\label{modified}
A(y)=ky+\frac{1}{\nu^2} \ln \left ( 1- \frac {y}{y_s} \right),
\eeq
where $k\sim M_{Pl}$ is the $AdS_5$ curvature, expected to be of the
order of the Planck mass scale, $y_s $ is  
the position of the metric singularity,
always chosen to be outside of the physical region, $y_s > y_1$, and $\nu > 0$ is a model parameter
 taken to be real. The $\nu$ parameter, alongside $\Delta = y_s -
 y_1$, the distance between the location of the metric singularity and
 the IR brane,  measures  the  departure of the metric from the pure
 $AdS_5$ background. The smaller the values of  $\Delta$ and $\nu$,
 the more modified the metric; intuitively, the singularity has a
 larger effect on the physics at the IR brane the  closer it gets to
 it. One can calculate the curvature along the 5-th dimension and obtain 
 \bea
 R(y)=8A''(y) - 20 \left(A'(y)\right)^2.
 \eea
 The curvature radius, $L(y) = \sqrt{-20/R}$, in units of $k$ along the 5-th dimension is then given by
 \bea\label{kLy}
 kL(y) = {k\Delta\over \sqrt{1-2\nu^2/5+2\nu^2k\Delta+(\nu^2k\Delta)^2}}.
 \eea
One can see that for values of $\nu>\sqrt{5/2}$, this function has a minimum 
before the singularity and therefore the curvature can change sign
within the physical region. Following \cite{Cabrer:2011fb},
we impose that this minimum is located outside of the physical region
and hence the curvature radius is a monotonically decreasing function
between the UV and the IR branes. 
 
 The more familiar RS metric is recovered by taking the limits $\nu
 \rightarrow \infty$ and $y_s \rightarrow \infty $, yielding
$A(y) =  ky$,
with the curvature radius being constant, $kL = 1$. 
\begin{figure}[!tb]
\bce
                \includegraphics[height=7cm]{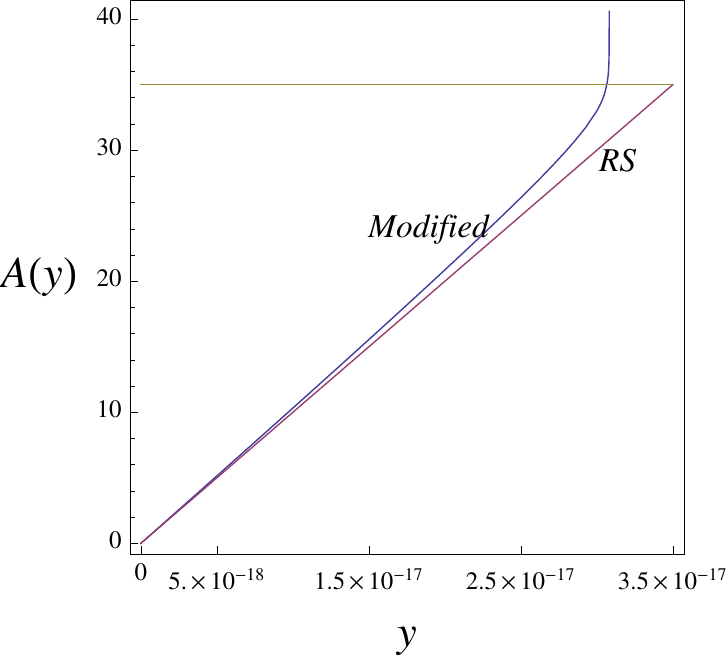}
\ece
\vspace{-.5cm}
\caption[The modified $AdS_5$ warp factor $A(y)$ versus the standard
  RS warp exponent, $y$]{The modified $AdS_5$ warp factor $A(y)$
  versus the standard RS warp exponent, $y$. The horizontal line
  corresponds to $ky=35$. For the same amount of warping, the modified scenario requires a
shorter length scale along the 5-th dimension. }
\label{fig:metric}
\vspace{.3cm}
\end{figure}
In Figure \ref{fig:metric}, we compare the two metrics and plot the warp exponent function $A(y)$
for the  $AdS_5$ and the modified $AdS_5$ cases. We can see that the
amount of warping near the IR brane at around  
$ky=35$ is larger for the modified $AdS_5$. Thus, as the figure indicates,  the same amount of
warping from the UV brane to the IR brane in the modified scenario requires a slightly smaller length of the
5-th dimension and hence an  IR brane slightly closer to the UV brane.
 The curvature radius (Eq. \ref{kLy}) at the UV brane is approximately equal for the pure and modified 
$AdS_5$ spaces with, $ky(y) \simeq 1$.  In contrary, at the IR brane, as $kL(y)$ is a monotonically 
decreasing function, $ky(y)$ assumes its minimal value for the modified $AdS_5$ space and hence
$kL_1\equiv kL(y_1)$ is a good measure of the amount of deviation from the pure $AdS_5$ space
with constant curvature radius. 

\begin{figure}[!ht]
\begin{center}$
\begin{array}{cc}
              \hspace{-1cm}
                \includegraphics[width=6cm,height=7cm]{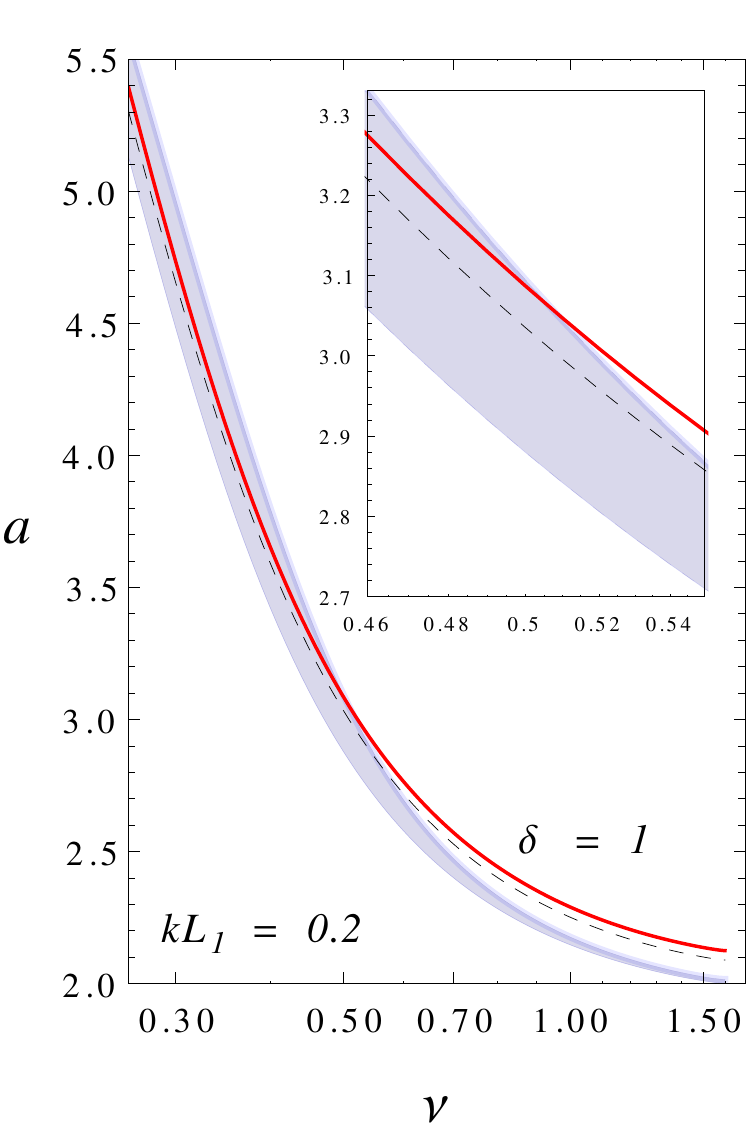}
               &\hspace{0.6cm}
               \includegraphics[width=6cm,height=7cm]{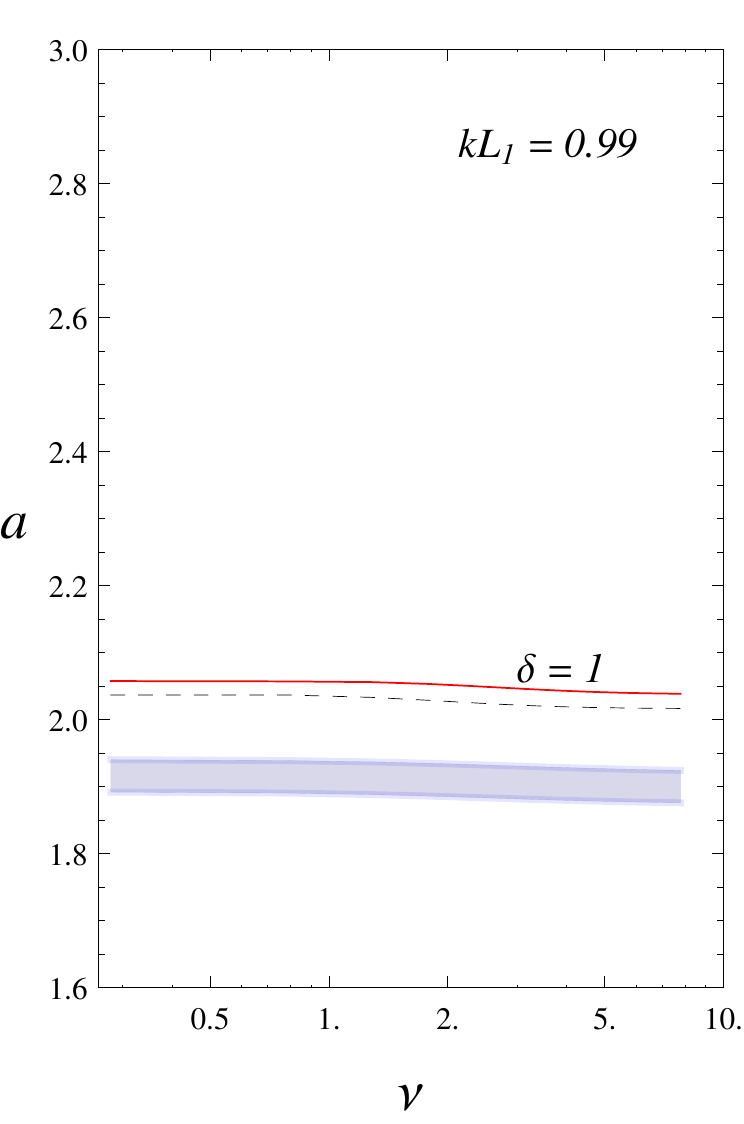} 
                \end{array}$
 \end{center} 
 \vskip -0.3in            
\caption{The $\delta=1$ plots and the connection with neutrino masses in the $\nu-a$ plane 
  for (left panel) $kL_1=0.2$ (large deviation) and (right panel) $kL_1=0.99$ (more
  RS-like). The red (solid) curve locates the no-fine-tuning threshold
   in which $\delta  = 1$. Above this curve, $\delta < 1$, and below the
   curve, $\delta > 1$ (corresponding to the unwanted fine-tuned
   region).  The shaded area  corresponds to a heavy neutrino mass
   ($m_3$ in {\it 
    normal} ordering), $m_{\nu}\sim5\times10^{-11}$~GeV for different
  values of the $c$-parameters, but such that the mass expression has no exponential
  sensitivity to the $c$-parameters. Note that in the RS-like case
  (right panel)  
  it is not possible to obtain neutrino masses without fine-tuning, or 
  quitting the neutrino plateau parameter region (as would be the case in usual
  RS scenarios). With modified-$AdS_5$ metric (left panel) it is
  possible to find non-tuned points with neutrino masses in the
  plateau (corresponding to no exponential $c$-dependence).
} \label{fig:aVSnu}
\end{figure}

The 5D fermion Lagrangian density with Dirac neutrinos is
\begin{eqnarray}\label{Lagrangian}
{\cal L}_q&=&
 {\cal L}_{kinetic} + M_{q_i}{\bar Q_i}Q_i + M_{u_i}{\bar U_i}U_i+ M_{d_i}{\bar D_i}D_i
 + (Y^{u\, 5D}_{ij} H\bar{Q}_iU_j + h.c.) \non \\
&&  + (Y^{d\, 5D}_{ij} H\bar{Q}_iD_j + h.c.)
+(Q_i\!\!\rightarrow\!\! L_i,  U_i\!\! \rightarrow\!\!N_i, D_i\!\! \rightarrow\!\!E_i)\, ,  
\end{eqnarray}
where, $i,j$ are flavor indices and the 5D Yukawa parameters, $Y_{ij}^{5D}$, are dimension-full quantities of
${\cal O}(1)\times \sqrt{k}$. $Q_i$ ($L_i$) are 5D quark (lepton) fields for $SU(2)$ doublets while $U_i$ ($N_i$)
and $D_i$ ($E_i$) are $SU(2)$ singlet quark (lepton) fields. The bulk
mass, $M_{\psi_i}$, originating from the momentum along the 5-th dimension,
can be taken in general to be $y$-dependent. To be able to compare, we
choose it such that it coincides with its usual definition in RS
models, and express  it in units of the 5-th dimension curvature, 
$k$, as $M_{\psi_i} = c_{\psi}^{i} k$, where $c_{\psi}^i$\footnote{ We use
  throughout $c_q$ for the doublets ($c_q$ and $c_l$) and $c_u$ for
  the singlets ($c_u$, $c_d$, $c_{\nu}$ and $c_e$).}  are
localization parameters, dimensionless quantities of ${\cal O}(1)$,
and $\psi_i$ runs over all SM quarks and leptons\footnote{Alternative fermion and Higgs profiles can be 
  found in \cite{Carmona:2011rd,Frank:2013qma}, where different bulk
  mass conventions are adopted.}. Dimensional
reduction then yields the normalized profile for the fermion and the
Higgs fields along the bulk of the extra dimension, $q^{0,i}_L(y)$,
$u^{0,i}_R(y)$ and $h(y) $, which are given by 
\bea
\label{q}
q^{0,i}_L(y)&=& q^i_0\ e^{(2-c_q^i) A(y)} \, , \\
\label{u}
u^{0,i}_R(y) &=& u^i_0\ e^{(c_u^i+2) A(y)}\, ,\\
\label{h}
h(y) &=& h_0\ e^{a k y}\, ,
\eea
with
$q^i_0 = f(c_q^i)  $,  
$\   u^i_0 = f(-c_u^i)  \ $ 
and 
$\  h_0=e^{- (a-1) k y_s} \mathfrak{h}_0$, 
and where $f(c)$ and $\mathfrak{h}_0$ are normalization factors which 
depend on $c$, $\nu$ and $y_s$,  given explicitly in Appendix \ref{app4}, along
with their limiting expressions for the usual RS ($AdS_5$) metric
background. From these profiles, one can
check that localization of the fields in the bulk of the extra dimension is determined 
by the values of the $c_{\psi}^i$ for the fermion fields, such that a value of $c_{\psi}^i > 1/2$ 
indicates a UV localized field, while a value of $c_{\psi}^i < 1/2$ localizes the field near the 
IR brane\footnote{This convention is  for left-handed doublets. For right-handed singlet fields, 
our convention is such that $c_{u}^i > -1/2$ for a UV localized field and $c_{u}^i < -1/2$ 
for an IR localized field.}. The Higgs field localization along the 5-th dimension 
is given by the parameter $a$, the dimension of the 
Higgs condensate operator. A completely IR brane localized field
corresponds to the limit $a\rightarrow\infty$, while for a delocalized Higgs field $a$ 
is small. However, in order to maintain the original Randall-Sundrum solution to the 
hierarchy problem without fine-tuning, the Higgs field localization
should be such that $a\gsim 2$ (for an $AdS_5$ metric background). 
If the 5D Higgs potential is of the form $V_{bulk}(H)= M_{5d}^2 H^2$, with
associated brane potentials at each boundary,   
the Higgs profile has two solutions, one growing towards the
IR and another one decaying at the IR brane. This last
one is proportional to $ e^{(4-a)ky}$ in the $AdS5$
background. In order to maintain the RS solution to the hierarchy 
problem, the decaying solution has be subdominant, and this happens
naturally for $a>2$. For $a<2$,  some fine-tuning between the
parameters of the bulk scalar potential and the brane potentials is
necessary in order to suppress the unwanted solution. In the
modified $AdS_5$ scenario the lowest value of $a$ that does not
require fine-tuning depends on the various new metric
parameters~\cite{Quiros:2013yaa}. In this case, the Higgs profile is given by
\bea\label{higgsfull}
h(y)
=h_0 e^{aky} \bigg[1 + (M_0/k -a) \left[ F(y) - F(0) \right]\bigg],
\eea
where $M_0$ is the brane Higgs mass term (coefficient of the $|H|^2\delta(y-y_1)$ term at the
IR brane) and the function $F(y)$ is given by
\bea
F(y) =  e^{-2(a-2) k y_s}k y_s \left[ -2(a-2) k y_s \right]^{-1 + 4/\nu^2} \Gamma \left[ 1 - \frac{4}{\nu^2} , -2(a-2) k( y_s - y) \right].
\eea
The decaying term at the IR brane is the second term in
Eq.~(\ref{higgsfull}) and,  
 forcing $M_0/k\simeq a$
(fine-tuning parameters),  the solution can become sub-dominant. In order to
avoid this fine-tuning of parameters, we note that as $F(y)$ is a monotonically
increasing function, and if one has $\delta \equiv |F(y_1)|\sim{\cal O}(1)$,  
no fine-tuning is needed to guarantee that the increasing solution for
the Higgs profile dominates. When the parameter $\delta=F(y_1)$
becomes larger, this solution needs fine-tuning of parameters to
suppress its value. 
Fig. \ref{fig:aVSnu} shows the no-fine-tuning
region (above the red solid $\delta = 1$ curve) in the $(a,\nu)$ plane, where $a$ is the
Higgs localization parameter and $\nu$ is the metric parameter of the
modified metric solution. The region below is the one fine-tuned, requiring an adjustment of Lagrangian parameters with a
tuning precision growing exponentially. (The close dashed curve
locates the points where $\delta = 10$, i.e. where the tuning is
already 10 $\%$) . 

In producing these graphs, for each
case we  first set the value of the IR brane position, $ky_1$,
which in turn fixes the value of $y_s$, the position of the
singularity. Then for each value of the parameter $\nu$ we solved for
$a$ in $\delta(a,\nu,y_1,y_s) = 1$.

To first order, the effective SM Yukawa couplings are obtained from the
overlap integral
\beq\label{YYY1}
y_{ij}^u = \frac{Y^u_{ij}}{\sqrt{k}} \int_0^{y_1} dy e^{-4A(y)} h(y) q^{0,i}_L(y)u^{0,j}_R(y)\, ,
\eeq
where the index $u$ denotes the four types of Yukawa couplings of
the SM, i.e. $u=u,d,e,N$ and we have defined the {\it dimensionless} 5D Yukawa couplings as 
$Y_{ij}^u = Y^{5D}_{ij}/\sqrt{k} \sim {\cal O}(1)$. Given the profiles (\ref{q}), (\ref{u}) and (\ref{h})
and the metric (\ref{modified}),  the integral above can be performed
analytically and written as
\bea\label{Y1}
y_{ij}^u = \tilde{Y}_{ij}^u h_0 f(c_q^i) f(-c_u^j),
\eea
where the factor $\tilde{Y_{ij}^u}$, defined in Appendix \ref{app4}, has very
mild $c^i$ dependence. The function $f(c)$ is such that depending on
the value of $c$, the Yukawa couplings can be
exponentially sensitive to $c$, or only mildly dependent. 
 Finally $h_0$ is the bulk Higgs normalization  factor and does not depend on the fermion mass
parameters $c^i$. 
Note that, although throughout the paper we have suppressed 
the explicit dependence of the fields on the metric
parameters, $\nu$, $y_s$, $y_1$ and $k$, and in particular all of the
factors of Eq.~(\ref{Y1}), are metric dependent. 
As shown in Appendix \ref{app4}, one can always retrieve the RS limit
for these terms by taking the limit $\nu, y_s \rightarrow \infty$.
\begin{figure}[!ht]
\begin{center}
      $
     \begin{array}{cc}
\hspace*{-1.1cm}
              \includegraphics[height=7cm]{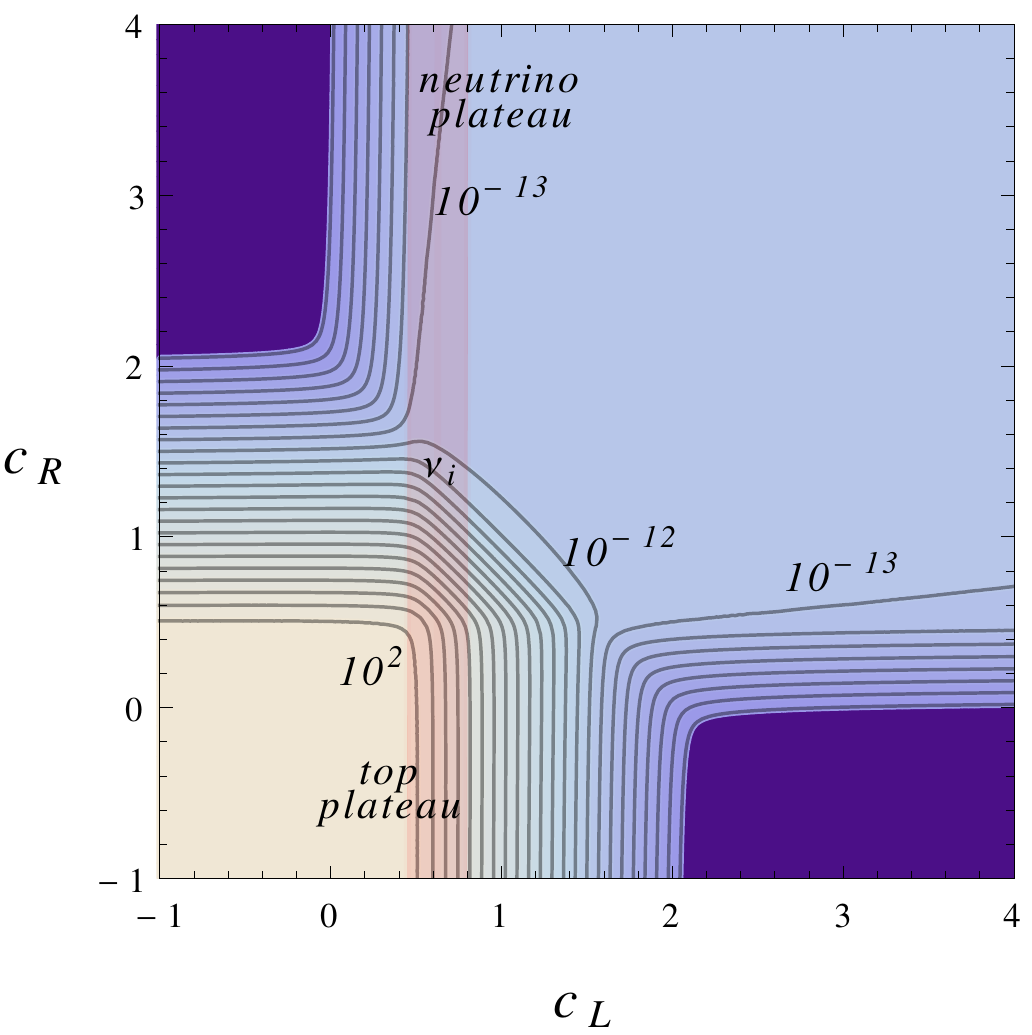}
              &\hspace{-0.1cm}
              \includegraphics[height=7cm]{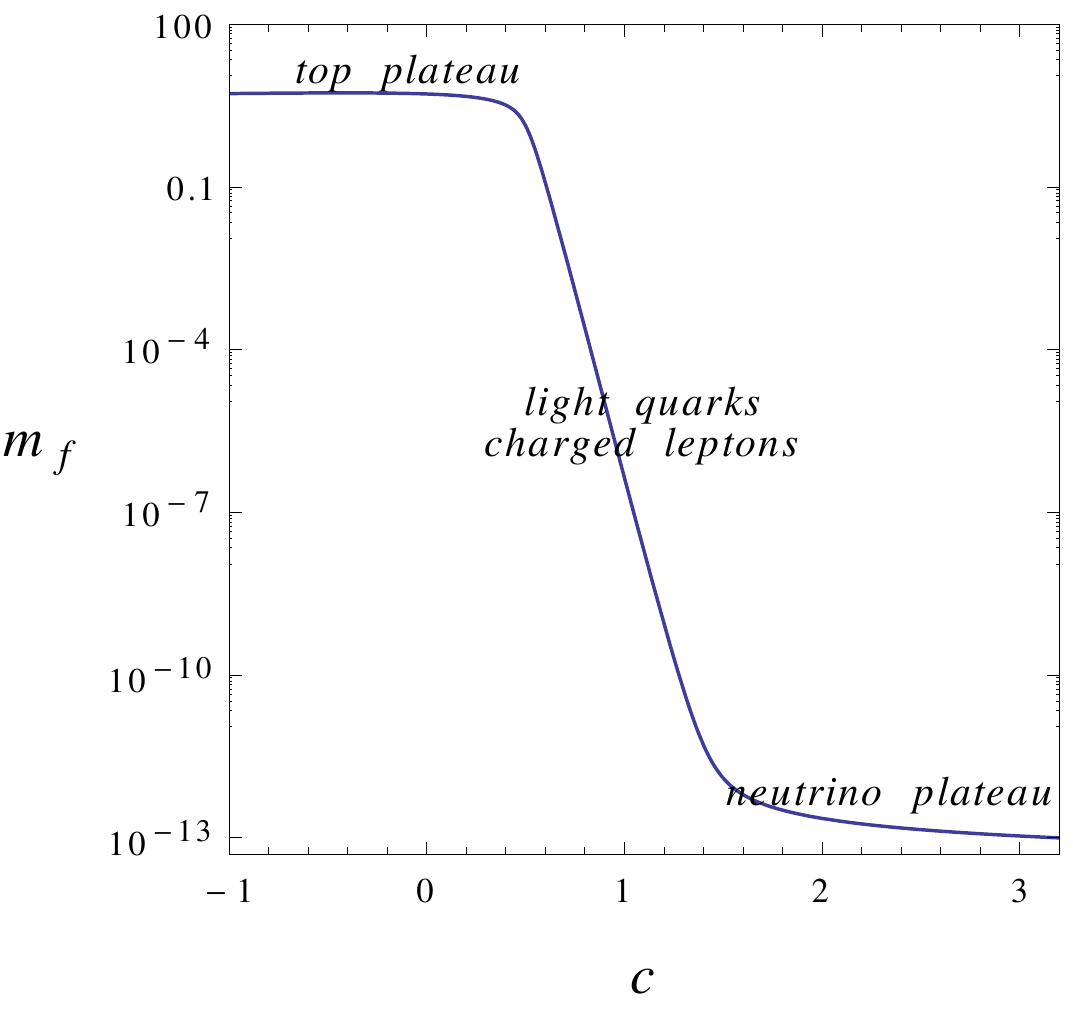} \\
      \end{array}
      $\\
                (a)  RS with all SM fields including the Higgs in the bulk, $a =2.04$ and $\delta = 1$. \\
\vspace{1cm}              
       $
       \begin{array}{cc}
\hspace*{-1.1cm}
              \includegraphics[height=7cm]{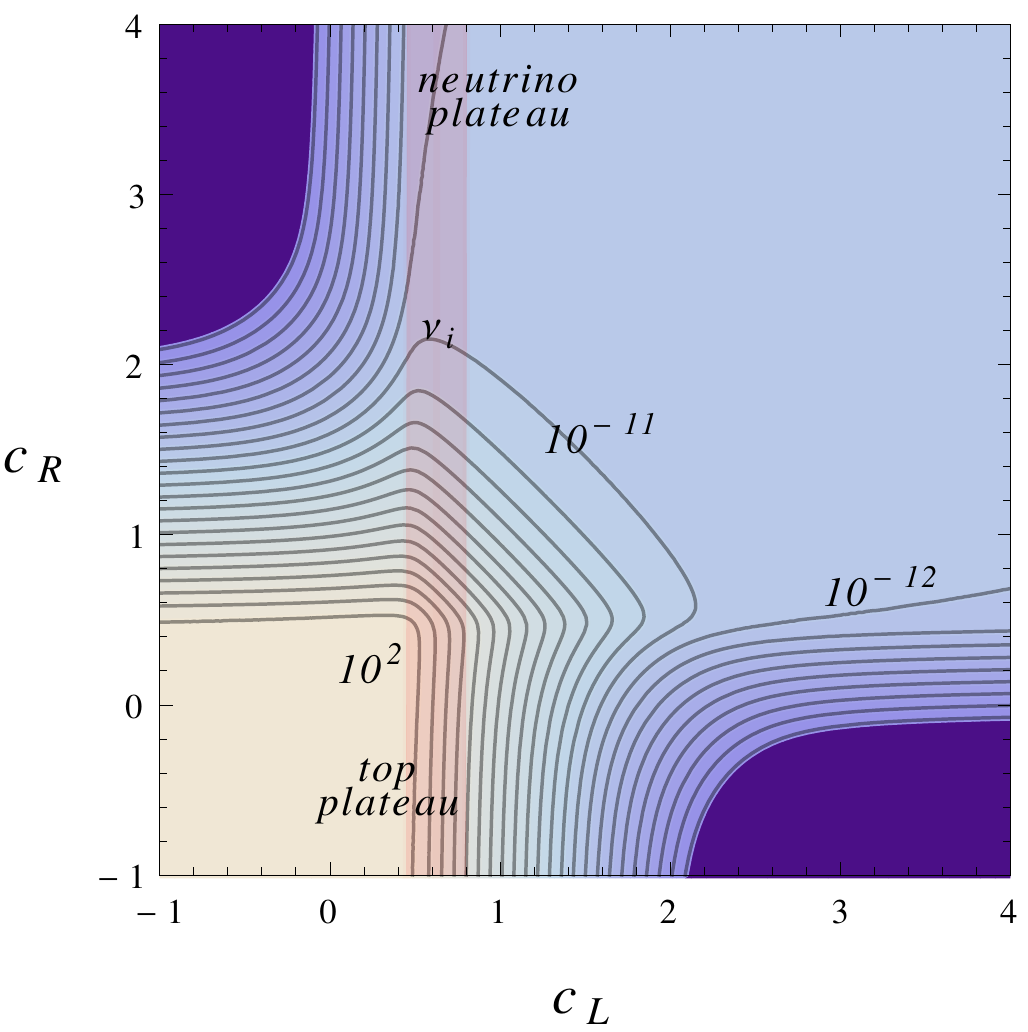}
                &\hspace{-0.1cm}
                \includegraphics[height=7cm]{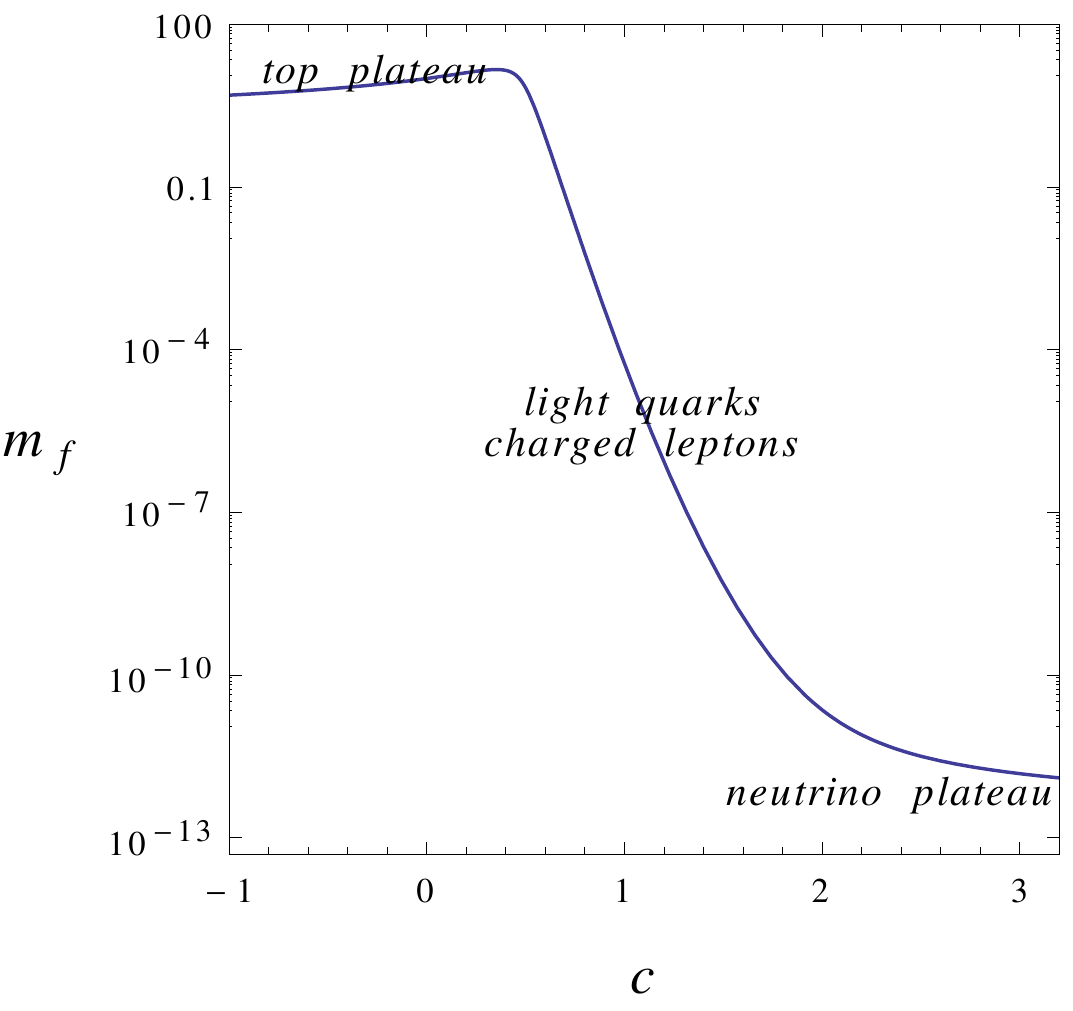}\\
         \end{array}
         $\\
                (b) Modified $AdS_5$ with $\nu=0.32$, $kL(y_1)=0.2$, $A(y_1)=35$, $a=4.46$ and $\delta = 1$.
\end{center}
\caption[Effective 4D Yukawa couplings for fermions]{Effective 4D
  Yukawa couplings for fermions as functions of the fermion bulk mass
  parameter $c$ for (a) the RS  and (b) general warped space-time metric. The plots in the right-hand side are produced by taking the
  $c$-parameters for the doublet and the singlet to be equal. In the
  left-handed plots, each contour depicts one order of  magnitude
  difference with respect to its adjacent contour. A typical location
  for neutrino masses of order $m_\nu\sim 10^{-11}$ GeV is also
  shown. The shaded vertical band shows the region where all fermions
  of the SM should be located.} 
\label{fig:c}
\end{figure}  

The fermion masses are given, to first order, by the eigenvalues of
the $3\times3$ Yukawa matrices of the form  
\bea
v\ y^{u} = v \left(\begin{array}{ccc} \label{Yij}
y_{11}^{u} & y_{12}^{u}  & y_{13}^{u}  \\
y_{21}^{u} & y_{22}^{u}  & y_{23}^{u} \\
y_{31}^{u} & y_{32}^{u}  & y_{33}^{u} \end{array}\right).
\eea
In general, in order to get the correct SM masses, no exponential
$c$-dependence is needed for the top quark, which corresponds to
$c_{q}^t \lsim 1/2$ and $c_{u}^t  > -1/2 $ (this region of parameter
space corresponds to the top plateau shown in Fig. \ref{fig:c}). For the rest of the
SM particles  $c_{q}^i \gsim 1/2$ and $c_{u}^i  < -1/2 $, which implies
that the Yukawa couplings will depend exponentially on the values of
the $c^i$.
In the case of neutrinos however, in order to accommodate  their tiny
masses, one needs localization parameters 
$c_{\nu} < -1 $. It was shown \cite{Agashe:2008fe} that, for a delocalized Higgs with $a$-parameter
small compared to the localization parameter $c^i$, the 4D effective
neutrino masses depend exponentially on $a$ but loose their dependence
on the $c^i$'s. This region of parameter space corresponds to the neutrino plateau shown in
Fig. \ref{fig:c}. In other words, the limit of the function in
Eq. (\ref{Y1}) for large $c_\nu$-s is given by
\bea
y_{ij}^u \sim {\tilde{Y}_{ij}^u } h_0.
\eea
To make this more transparent, in the formula for fermion masses given by (\cf ~ Eq.(\ref{Y1}))
\bea\label{Mf1}
(M_f)_{ij} = v \tilde{Y}_{ij}^u h_0 f(c_q^i) f(-c_u^j),
\eea
we factor the exponential behaviors and write them in the following two distinct limits
\begin{eqnarray}\label{Mf}
(M_f)_{ij} &   \sim &
  v Y_{ij}^f \epsilon^{(c_q^i-\frac{1}{2})}\eps^{-(c_u^j+\frac{1}{2})}
  \ \ \text{for} \ \ \ \ \ \ \ c_q>1/2, c_u <-1/2 \, ,\nonumber\\ 
(M_\nu)_{ij}& 
  \sim & v Y_{ij}^N
  e^{-ky_1(a-1)}\ \ \ \, \  \ \ \    \text{for}
  \ \ \ \ \ \ \ \ \ \ \ c_q - c_u > a\, , 
\end{eqnarray}
where $\eps = e^{-A(y)}$ (see Appendix \ref{app4} for details)\footnote{For modified $AdS_5$, $\eps_{mod} = e^{-k y_1} \left( 1-\frac {y_1}{y_s} \right)^{\frac{1}{\nu ^2}} \sim 10^{-15}$ while for RS 
$\eps_{RS} = e^{-k y_1} \sim 10^{-15}$.}. 

The limits described in Eq. (\ref{Mf}) imply
that, as the quark and charged lepton mass matrix elements are
exponentially dependent on the $c^i$ parameters, any structure in the
5D Yukawa matrix elements $Y^f_{ij}$ will be largely washed-out and
will always produce generically hierarchical fermion masses as well as small 
mixing angles. For the neutrinos, on the other hand, since there is no
exponential sensitivity on the flavorful $c^i$ parameters, any
structure inherent in the 5D Yukawa matrix elements will survive in
the 4D effective theory. This is the region of neutrino parameter space we are
interested in, shown in Fig. \ref{fig:c} as the neutrino plateau.   
In RS models, the actual height of the neutrino plateau is determined exclusively 
by the value of the $a$ parameter. For the value of the warp factor required to solve the hierarchy
problem, the highest possible neutrino masses in the plateau are too small by 1-2 orders of magnitude (without fine-tuning) to be
phenomenologically viable  (see upper panels in Fig. \ref{fig:c}). 
Some  tuning, or some enhancement of the 5D Yukawa couplings, and/or trespassing the edge of the
plateau would then be required, making the scenario less attractive
for our purposes.  
 On the other hand, in modified $AdS_5$ scenarios, although the level 
of the plateau is still highly sensitive to the value of $a$, the masses could
actually be increased by some two orders of magnitude and thus
allow for  
 phenomenologically acceptable neutrino masses in the desired region
of the model (see Fig.~\ref{fig:aVSnu}  and lower panels in
Fig. \ref{fig:c}).
This feature occurs because the modified $AdS_5$ metric (see Eq. (\ref{modified}))  
exhibits a richer parameter space. In the RS metric, in order to 
produce the correct neutrino masses in the plateau, we need
 $a\sim 1.85$,  value which amounts to about $1/10000$ fine tuning of
parameters in the 5D Higgs potential. While for modified $AdS_5$ scenarios,
 one can produce a neutrino plateau within the experimental
bounds for some range of $a$ parameter values without fine-tuning. 
It is interesting that the parameter space for which the neutrino plateau is most favorable coincides with the
region where small KK masses satisfy bounds coming from flavor and electroweak precision
measurements \cite{Cabrer:2011qb,deBlas:2012qf} {\it and} constraints from Higgs 
phenomenology \cite{Frank:2013un}.

\begin{figure}[!htb]
\bce$
\begin{array}{cc}
              \hspace{-1.5cm}
                \includegraphics[width=7cm,height=7cm]{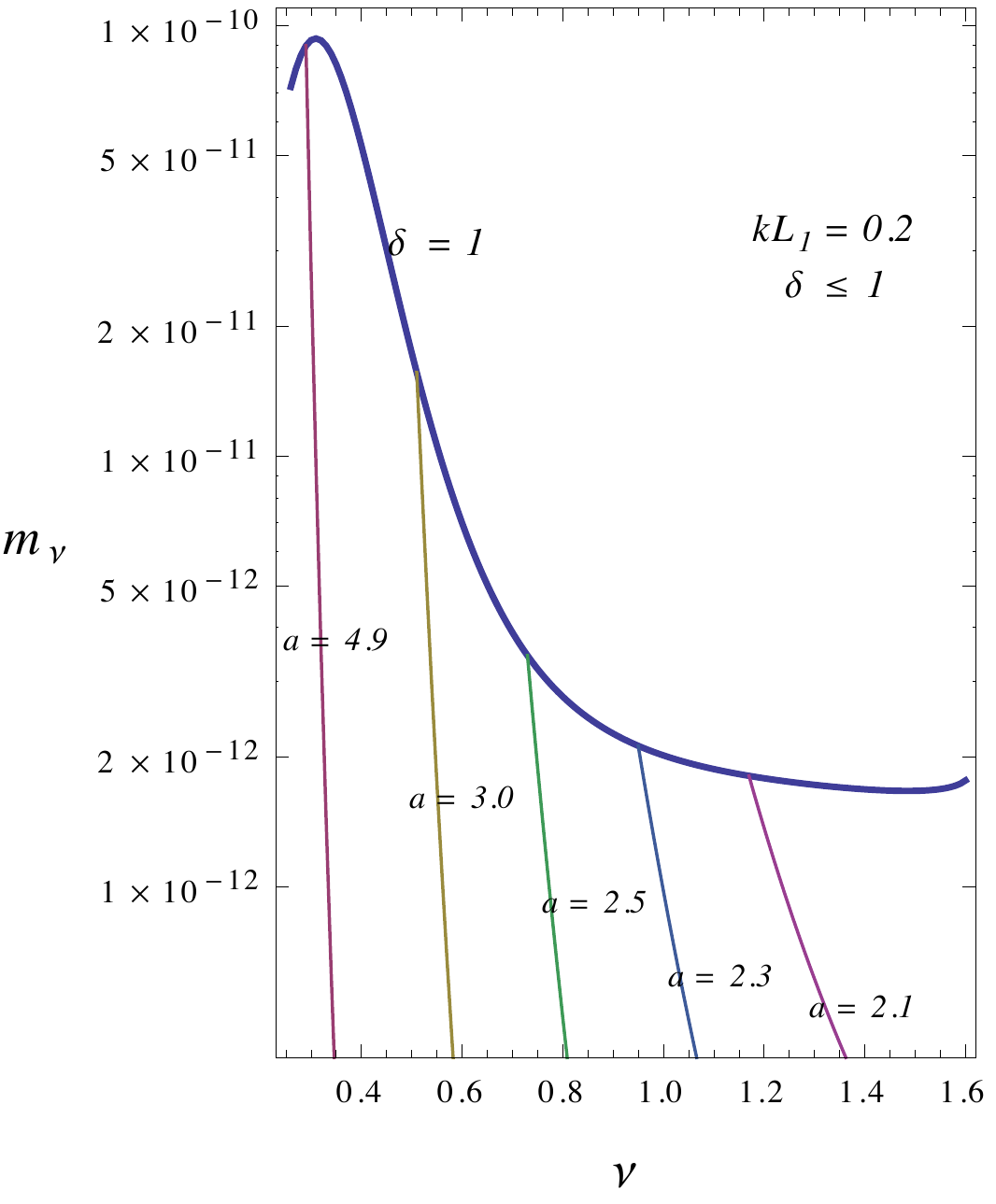}
                  &\hspace{-0.3cm}
               \includegraphics[width=7cm,height=7cm]{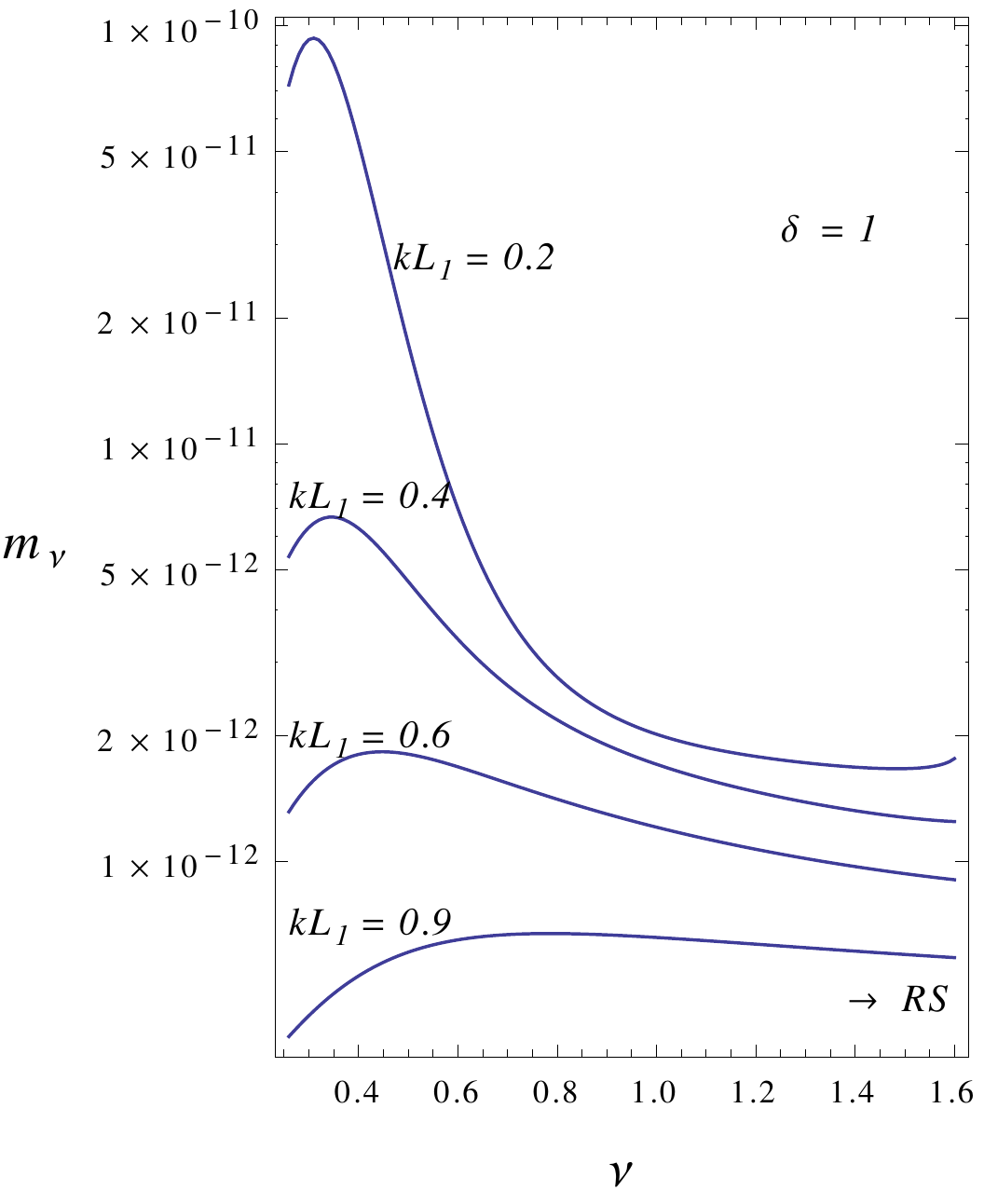} 
                \end{array}$
\ece
\vskip -0.3in
\caption[The neutrino masses versus the $\nu$ parameter.]{(Left panel) Neutrino
  masses as  functions of the metric parameter $\nu$ for different
  values of the Higgs localization parameter $a$ and fixed $kL_1=0.2$. The fermion mass
  parameters ($c$'s) are fixed to a region where
  there is no exponential dependence on them (the neutrino
  plateau). The curves end  whenever the fine-tuning threshold
  ($\delta =1$) is reached (thick   overlapping curve). Note that for
  small values of $\nu$ the neutrino masses become larger while
  still remaining in the non-tuned regime and in the neutrino plateau.  
  (Right panel) Same plot for different values of 
  $kL_1$ and fixed $\delta = 1$. 
  The graph corresponding to $kL_1=0.9$ remains constant at
  larger values of $\nu$ and (approximately) coincides
  with the RS limit.}  
\label{fig:muVSnu}
\end{figure}

In order to further illustrate this issue,  in
Fig.~\ref{fig:muVSnu} we show the resulting neutrino masses in the plateau
region as a function of the parameter $\nu$,  for different values
of $a$ (left panel) and $kL_1$ (right panel). The area below the curves is the no-fine-tuning region,
and one can see that the largest neutrino masses (in the plateau)
occur for $\nu\sim 0.3$ and $kL_1=0.2$, whereas in the RS limit,
{\it i.e.} $kL_1\sim 1$ and $\nu$ large, the masses are some two orders of
magnitude lower (slightly too low). This makes the extended metric scenarios a more natural
framework for the flavor mechanism investigated here, and adds to the
advantages of these scenarios ({\it i.e.} a much lower KK scale
consistent with electroweak and flavor bounds, and with Higgs production). 

Qualitatively the general features of the flavor
structure of these modified $AdS_5$ models are very similar to these
features in the pure $AdS_5$ models for the bulk mass $c_\psi^i$
parameters. Therefore, in order illustrate our flavor setup,  it is
 useful to consider the simpler case of the RS metric. In 
this case the fermion mass formulas Eq. (\ref{Mf}), 
can be simplified dramatically. As usual, we consider the mass matrix
for the neutrino sector separately from the case of quarks and charged
leptons mass matrices. Consider the case with $c_{q,u}^i >
\frac{1}{2}$. In this case we have,
\begin{eqnarray}\label{RSMass}
(M_f)_{ij} &\simeq& v\epsilon^{(c_q^i-\frac{1}{2})}\eps^{-(c_u^j+\frac{1}{2})} \sqrt{2(a-1)|1-2c_q^i||1+2c_u^j|}\tilde{Y}_{ij}^f
\nonumber\\
(M_\nu)_{ij}&\simeq& v\epsilon^{a-1}\frac{\sqrt{2(a-1)|1-2c_l^i||1+2c_{\nu}^j|}}{\sqrt{\epsilon^{(1-2c_l)}-1)}
\sqrt{\epsilon^{(1+2c_\nu)}-1} }\tilde{Y}_{ij}^{\nu},
\end{eqnarray}
where the 5D Yukawa couplings are given by\footnote{For exact formulas see Appendix \ref{app4}.},
\bea
\tilde{Y}_{ij}^u \simeq \frac{1}{a-c_q^i+c_u^j}Y_{ij}^u.
\end{eqnarray}
From Eq. (\ref{q}), (\ref{u}) and (\ref{YYY1}) one can see that there are two sources of flavor structure: 
the 5D dimensionless Yukawa couplings 
$Y^{u}_{ij}$, and the bulk mass coefficients $c_{\psi}^i$. We are
interested in scenarios in which all Yukawa matrices ($Y_F = $
$Y^{u}_{ij}$, $Y^{d}_{ij}$, $Y^{\nu}_{ij}$, and $Y^{e}_{ij}$) and
fermion bulk mass matrices from the 5D Lagrangian (${\mathbf c}_f = $
${\mathbf c}_{q}$, ${\mathbf c}_{u}$, ${\mathbf c}_{d}$, ${\mathbf
  c}_{l}$,  ${\mathbf c}_{\nu}$, and ${\mathbf c}_{e}$) share the same
symmetry structure, which is then slightly broken through some high
energy mechanism according to 
\bea\label{Yexpansion}
Y_F = Y^{0}_{F} + \delta Y_F,
\eea
\bea\label{cexpansion}
{\mathbf c}_f = {\mathbf c}_f^0 + \delta{\mathbf c}_f,
\eea
where the matrices $Y_F^0$ and ${\mathbf c}_f^0$ are flavor symmetric while the
perturbation matrices $\delta Y_F$ and $ \delta{\mathbf c}_f$ are random.
Inserting these perturbations in Eqs. (\ref{RSMass}),  the fermion masses receive corrections to  leading order as follows
\bea\label{MpertRS}\nonumber
m_{t}\   &=& \ m^0_{t} +\delta m_{t}  \hspace{6.8cm} c_{q}^3 , c_{u}^3 <1/2 \, , \\
(m_f)_{ij} &=& (m_f)^0_{ij}\  f(\delta c_q^i) f( \delta c_u^j)) \sim (m_f)^0_{ij}\  \epsilon^{( \delta c_q^i+ \delta c_u^j)}  
\hspace{1.6cm} a > c_l^i+c_u^j \, ,\\
(m_\nu)_{ij}&=& (m_\nu)^0_{ij} + \delta (m_\nu)_{ij}  \hspace{5.7cm} a< c_{l}^i+c_{\nu}^j\ \nonumber\label{neutrinoM} .
\eea
Therefore the same exponential sensitivity on the bulk mass $c_\psi^i$ parameters, $\epsilon\sim10^{-15}$, 
responsible for producing the SM hierarchy in the standard RS, is now translated into exponential sensitivity of the
symmetry breaking terms. As a consequence, small symmetry
breaking terms ($|\delta c^i| \sim 0.1 $) can
produce mass corrections of order $10^{-15(\delta c^i +\delta c^j)}$ (\ie
~a hierarchy of order $\sim10^{6}$) to the quark and charged lepton
mass matrices. This is in complete agreement with the observed
hierarchy in these sectors. As mentioned before, the neutrinos and the
top quark fields live in the two plateaus (see Fig. \ref{fig:c})
with mild $c_\psi^i$ sensitivity. 


For the mixing angles, the eigenvectors matrix that
diagonalizes the neutrino sector should be very close to the eigenvectors
matrix of the 5D Yukawa matrix. However, in the quark and charged lepton
sectors the mixing matrices are generically close to the unit matrix
with off-diagonal entries hierarchically small as\footnote{Similar
  expressions are obtained for $V_L^d$ and $V_L^q$ (see for example \cite{Casagrande:2008hr}).}:
\begin{equation}\label{Vu11}
V_L^{u} \simeq \left(\begin{array}{ccc} 1 & \frac{f^1_q(\tilde{M_{u}})_{21}}{f^2_q(\tilde{M_{u}})_{11}}
& \frac{f^1_q(\tilde{Y_{u}})_{13}}{f^3_q(\tilde{Y_{u}})_{33}}\\
-\frac{f^1_q(\tilde{M_{u}^*})_{21}}{f^2_q(\tilde{M_{u}^*})_{11}}& 1 & \frac{f^2_q(\tilde{Y_{u}})_{23}}{f^3_q(\tilde{Y_{u}})_{33}}\\
\frac{f^1_q(\tilde{M_{u}^*})_{31}}{f^3_q(\tilde{M_{u}^*})_{11}}& -\frac{f^2_q(\tilde{Y_{u}^*})_{23}}{f^3_q(\tilde{Y_{u}^*})_{33}}& 1 \end{array}\right),
\end{equation}
where $f_q^i$ is shorthand for the profile functions $f(c_q^i)$, $(\tilde{M_{u}})_{ij}$ is 
the $(ij)$ minor of the matrix in parenthesis, and $(\tilde{Y}_u)_{ij}
$ is the $ij$ element of the Yukawa matrix.
 We define the CKM and the PMNS matrices as the following
\bea\label{pdg2}
V_{CKM} \equiv V_L^u V_L^{d\dagger} \ \ \ \ \ \ \ \ \ \ \text {and
}\  \ \ \ \ \ \ \   \ V_{PMNS} \equiv V_L^\nu V_L^{e \dagger}\ . 
\eea
As mentioned after Eq. (\ref{Mf}), for the quarks and charged leptons, the off-diagonal mixing angles are
also exponentially sensitive to the $c_\psi^i$ parameters and hence
to the small symmetry breaking terms. The $V^f_L$ matrix elements 
 in Eq. (\ref{Vu11}) can then be written as
\bea\label{lepquarkmixing}
(V_{L}^{u,d,e})_{ij}  
\sim (V_{L}^{u,d,e})_{ij}^0\  \epsilon^{(\delta c_{q}^i - \delta c_q^j)} \ \ \ \ \ \ \ \ \  \ \ \ \ \ \text{(\ for u, d and e) }\, ,
\eea
where  $(V_{L}^{u,d,e})_{ij}^0$ are the matrix elements before the
flavor symmetry breaking and $\{i,j\}$ can be $\{1,2\}$, $\{1,3\}$ and $\{2,3\}$.
We see again that, due to  exponential warping, all original symmetries present in the high energy theory
are washed out in the quark and charged lepton sectors.
Contrary to this, in the Dirac neutrino sector,  the terms in the
$V_L^\nu$ are much less sensitive to the symmetry breaking terms, since
their own dependence on the $c_\psi^i$ is mild in the region of parameters
of interest. If we define the eigenvalues of the 5D neutrino Yukawa 
matrix as
\bea
Y_\nu^{diag} = V_{Y_L} Y_\nu V_{Y_R},
\eea
then the matrix diagonalizing the 4D effective neutrino mass matrix
\bea
(V_{L}^{\nu})_{ij} \sim (V_{Y_{L}})_{ij} \ .
\eea
 So far the
framework is general, and we did not specify  the type of symmetry imposed. In the next two sections we will consider two concrete implementations, one within flavor complementarity and then a second within  flavor democracy. The background metric considered will
always be the modified $AdS_5$ solution, in the most favorable region of
parameter space.

\section{Flavor complementarity}
\label{sec:02}
In the first symmetry implementation, in this section we study the effects of enforcing a strong
correlation between 5D quark and 5D lepton
parameters.  
This scenario is motivated by the observation that charged
lepton masses and down quark masses obey similar patterns. In
warped space models, masses are obtained from exponentially small 
overlap integrals, and having similar mass patterns can indicate that
both the 5D Yukawa structure and the 5D bulk fermion
masses ($c$-parameters) are very similar for the down quarks  and for
the charged leptons. We therefore 
assume that $Y_d = Y_e$ and
${\bf c}_q ={\bf  c}_l$ to a high order of precision, so that the
differences in observed masses between down quarks and charged leptons
come from deviations in the right handed bulk parameters ${\bf c}_d$
and ${\bf  c}_e$. 

At the same time,  
inspired by neutrino phenomenology, we l 
implement a popular
 2-3 ($\mu-\tau$) symmetry 
\cite{Ma:2001mr,Grimus:2001ex,Mohapatra:2004hta,Mohapatra:2005ra,Nasri:2005gw,Mohapatra:2006un,Mohapatra:2006pu,Xing:2006xa,Ahn:2006nu,Aizawa:2005yy,Fuki:2006ag,Joshipura:2005vy,Grimus:2009pg,Adhikary:2009kz,Xing:2010ez,Ge:2010js,deMedeirosVarzielas:2011tp,He:2011kn,Gupta:2013it,Adhikary:2013bma,Adhikary:2012mt,Liao:2012xm,Hamzaoui:2013twa,Lashin:2013xha}
within the general flavor structure of the model. We  
 require that all the 5D Yukawa matrices
must be symmetric under 2-3 permutations and consider, for simplicity, the case where the 5D fermion bulk mass matrices ${\bf c}_f^0$
are diagonal and degenerate in that basis (maybe a remnant of some global
$U(3)$ flavor symmetry broken in the Yukawa sector), i.e.  
\bea
{\bf c}^{0}_{f}=\!\! \left(\begin{array}{ccc}
 {c^{\,0}_{f}} & 0 & 0 \\
0 & c^{\,0}_{f}  &  0\\
0 & 0 & c^{\,0}_{f} \end{array}\right),\ \ \ \ \ \
\eea
where $c^{0}_{f}$ are real parameters, and $f$ denotes both
doublets and singlets $f=q,l,u,d,e,\nu$. 
In 4D scenarios, discrete permutation symmetries  
of the neutrino mass matrix are
known to lead to interesting (and phenomenology viable) neutrino
masses and mixing patterns (such as the tri-bimaximal mixing matrix
\cite{Harrison:1999cf,Harrison:2002er,Xing:2002sw,Harrison:2002kp,Harrison:2003aw,He:2003rm}, or the bi-maximal mixing matrix \cite{Vissani:1997pa,Barger:1998ta,Baltz:1998ey,Stancu:1999ct,Georgi:1998bf,Li:2004nh}).

We will impose the 0-th order Yukawa matrices to be 
\bea\label{23Y}
Y^{0}_{u,\nu} =y_u \left(\begin{array}{ccc}
\sqrt{6}d & 0 & 0 \\
d & d+1 & d-1\\
d & d-1 & d+1\end{array}\right)   \ \  {\rm and}\ \ \ \
Y^{0}_{d,e}= y_d \left(\begin{array}{ccc}
 a & 1 & 1 \\
c & g  & e\\
c & e & g \end{array}\right),\ \ \ \ \ \
\eea
where the parameters $y_u$, $y_d$, $a$, $c$, $d$, $g$ and $e$ are
complex.\footnote{We  
also assume that the symmetry is broken by some 
small perturbations, i.e.  $Y_i =Y^0_i + \delta Y_i$ and ${\bf c}_{f}=  {\bf c}_{f}^0+\delta {\bf c}_{f}$,
although we  
require that the constraints  $Y_d = Y_e$ and
${\bf c}_q ={\bf  c}_l$ survive the flavor symmetry breaking in
order to maintain the quark lepton complementarity.}
 As explained above, we  
 impose the down type Yukawa couplings to match the charged
lepton ones, i.e. $Y_d = Y_e\equiv Y_{d,e}$. We further require that
the up-type Yukawa couplings be the same as the neutrino 
Yukawa couplings, i.e  $Y_u = Y_\nu\equiv Y_{u,\nu}$, in order to
reduce free parameters and enhance the degree of correlation between quarks and leptons.

The structure of the down-type Yukawa $Y^0_{d,e}$ is set only by the
2-3 symmetry, but the structure of neutrino-type Yukawa couplings $Y^0_{u,\nu}$ in Eq.~(\ref{23Y}) requires more
explanation. The number of parameters has been reduced to two complex
parameters, $y_u$ and $d$, such that the rotation matrix that diagonalizes $Y^0_\nu$ is
the bi-maximal mixing matrix\footnote{The tri-bimaximal scheme predicts the CP phase in the PMNS matrix, $\delta$, to be zero, contrary to phenomenological conjectures which favor $\delta=-\frac{\pi}{2}$.} and the eigenvalues of $Y^0_{u,\nu}$ are  
simply given by $3.08\ d $, $1.59\ d  $ and $2$ (in
units of $y_u$).  When $|d| \sim 1/6$,  we should naively generate a hierarchy between the
eigenvalues of the matrix matching the solar and atmospheric neutrino
mass hierarchies, for the case of a normal ordering in the neutrino masses.  
Of course, in warped scenarios, the 5D Yukawa matrices are  in
general not directly proportional to the effective 4D fermion mass matrices. In the
neutrino sector however, since we are near the neutrino mass ``plateau'' of
the $c$-parameters, the structure of the effective 4D mass matrix should
be very similar to the 5D Yukawa 
matrix. 
 We therefore expect that the effective 4D neutrino
structure should be diagonalized by a matrix close to
the bi-maximal mixing matrix, with a normal ordering in the masses
matching the observed mass differences.  5D wave function
effects, as well as random, but small, symmetry breaking terms, would
perturb this structure, but the main features  
survive. 

Finally, the two texture zeroes of $Y^0_u$ in Eq.~(\ref{23Y}) cause the vanishing of the (21) minor  of that matrix, which  
in turn contributes to the quark-lepton correlations which we want
to address here; the vanishing, or the smallness of those entries, is
therefore  a critical requirement in the setup, as  
it  
links the value of the Cabibbo
angle to the PMNS element $V_{13}$ (i.e. to $\theta_{13}$). To see how
 occurs, let's write the approximate
expression for the Cabibbo angle $V_{us}$ in warped extra dimensions,
\bea
V_{us} \simeq  \left| {f(c^1_{q})\over f(c^2_{q})}\right| \left| \frac{(\tilde{M_{d}})_{21}}{(\tilde{M_{d}})_{11}} - \frac{(\tilde{M_{u}})_{21}}{(\tilde{M_{u}})_{11}}\right|,
\eea
where $\tilde{(M_{u}})_{ij}$ and  $\tilde{(M_{d}})_{ij}$ are the  
$(ij)$ minors of the $Y_u$ and  $Y_d$ matrices. We can see that the texture
zeros in the 5D Yukawa matrix $Y_u$ ensure that $\tilde{(M_{u}})_{12}
= 0$, so that the Cabibbo angle is controlled, to 
first order, by the down sector Yukawa couplings and by the quark
doublet bulk $c$-parameters, i.e. 
\bea\label{Cabb1}
V_{us} \simeq   {f(c^1_{q})\over f(c^2_{q})} \frac{(\tilde{M_{d}})_{21}}{(\tilde{M_{d}})_{11}} \, . 
\eea
Since we imposed $Y_d^0 = Y_e^0$ and ${\bf c}_q = {\bf c}_l$, the left mixing
matrices $W^d_L$ and $W_L^e$ 
for the down sector and charged lepton sector  
 are very similar, and in particular the 12 entries  
 have the exact same
first order expansion $(W_L^d)_{12}\simeq (W_L^e)_{12} \simeq
V_{us}$. This is the origin of the quark-lepton complementarity relations we
wish to study here and to check their robustness against flavor symmetry
perturbations.

In the lepton sector, the effective 4D neutrino mass matrix should be diagonalized by a unitary
matrix, almost bi-maximal, with its third eigenvector (associated to $m^\nu_3$) close to
$(0,-\frac{1}{\sqrt{2}},\frac{1}{\sqrt{2}})$. The charged lepton
Yukawa coupling, on the other hand, will be diagonalized by $W^{d}_L$, since both
the Yukawa couplings and the doublet $c$-parameters are the same in the down quark and charged
lepton sectors. The unitary matrices
generating the charged flavor part of the PMNS matrix $V_{PMNS}=
(W^{e}_L)^{\dagger} W^{\nu}_L$ should be close to 
\bea
W^{e}_L\simeq \left(\begin{array}{ccc} 1 & V_{us} & V_{ub} \\
-V_{us}^* & 1 & \frac{V_{ub}}{V_{us}}- {f(c^2_{q})\over f(c^3_{q})}  \\
- {f(c^2_{q})\over f(c^3_{q})} V_{us}^* & -\frac{V_{ub}^*}{V_{us}^*}+
{f(c^2_{q})\over f(c^3_{q})}  & 1 \end{array}\right)
{\rm\ \   and\ \ \ }
 W^{\nu}_L\simeq \left(\begin{array}{ccc} 
\frac{1}{\sqrt{2}}  &   \frac{1}{\sqrt{2}} &  0\\
- \frac{1}{2}       &  \frac{1}{2}         & -\frac{1}{\sqrt{2}}  \\
- \frac{1}{2}      &   \frac{1}{2}         &  \frac{1}{\sqrt{2}}
\end{array}\right) + {\cal O }(\epsilon)\, ,\ \ \ \ \
\end{eqnarray}
where we have $W^d_L \simeq W^{e}_L$, and where the ${\cal O }(\epsilon)$ entries represent small
corrections coming from wave-function effects and flavor symmetry breaking terms.
The PMNS matrix elements
$V_{e3}$, $V_{\mu 3}$ and $V_{e2}$ are then expected to be 
\begin{eqnarray}
V_{e3} &\simeq&  \left(1-{f(c^2_{q}) \over f(c^3_{q})} \right) \frac{V_{us}}{\sqrt{2}}+ {\cal O }(\epsilon) \label{e3us}  \, ,\\
V_{\mu 3} &\simeq& -\frac{1}{\sqrt{2}}\left[1 +\frac{V_{ub}}{V_{us}}-{f(c^2_{q})\over
    f(c^3_{q})}\right] + {\cal O }(\epsilon) \, ,  \\
V_{e2} &\simeq&
\frac{1}{\sqrt{2}} -\frac{1}{2} \left(1 + {f(c^2_{q})\over f(c^3_{q})}  \right) V_{us} + {\cal O }(\epsilon) \, .
\eea
\begin{figure}[t]
\begin{center}
               \includegraphics[height=2in]{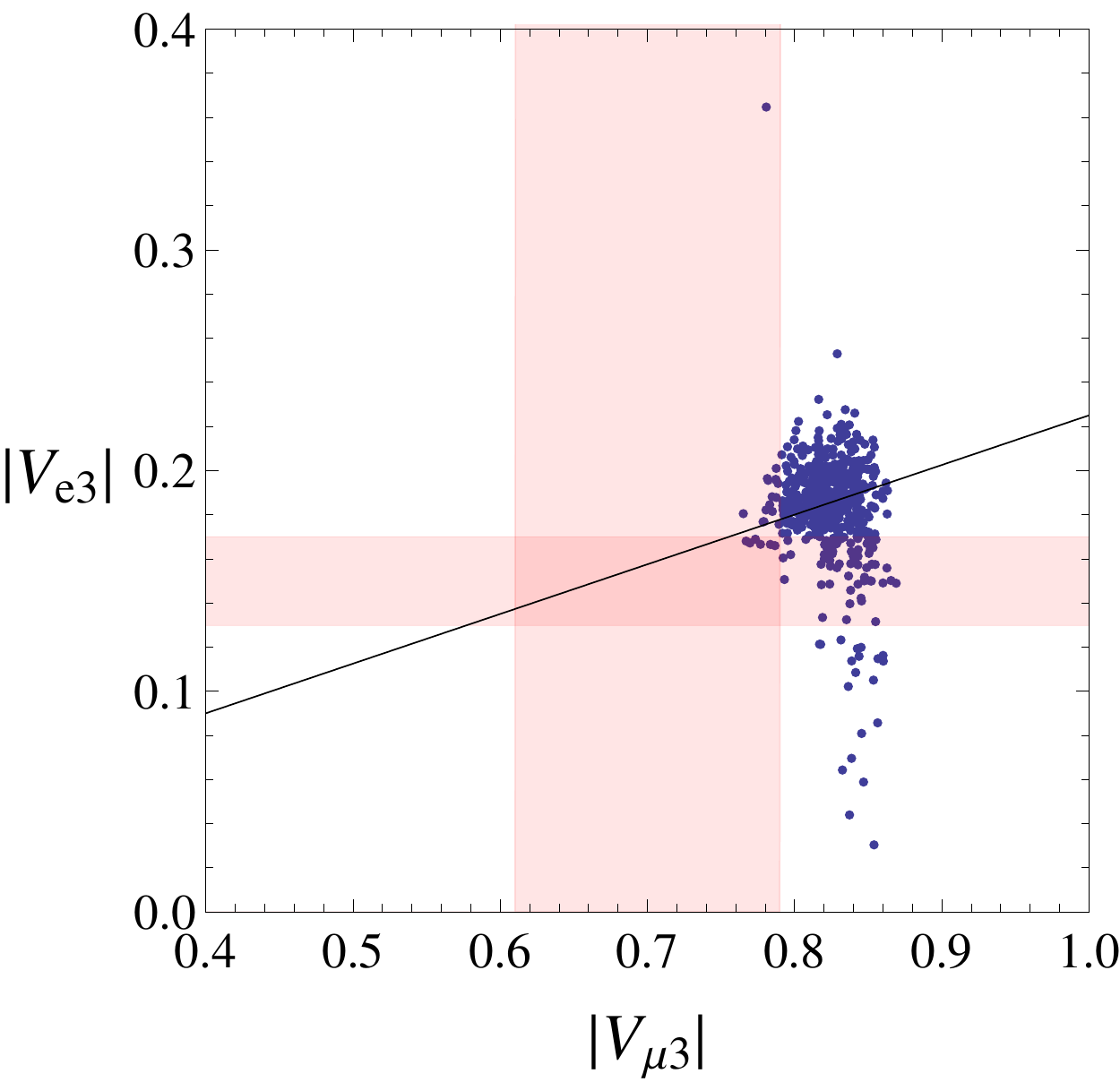}
               \includegraphics[height=2in]{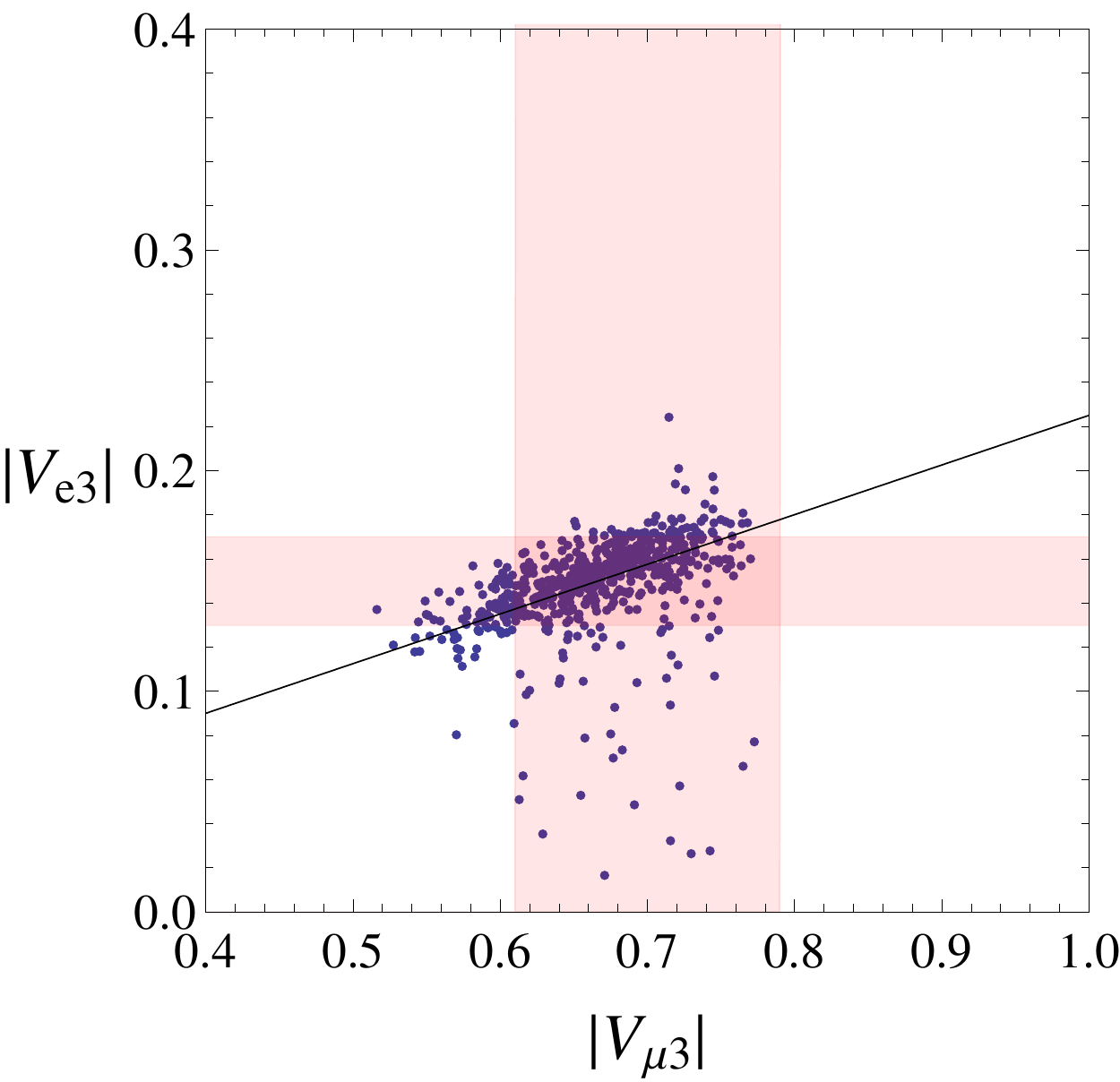} 
               \includegraphics[height=2in]{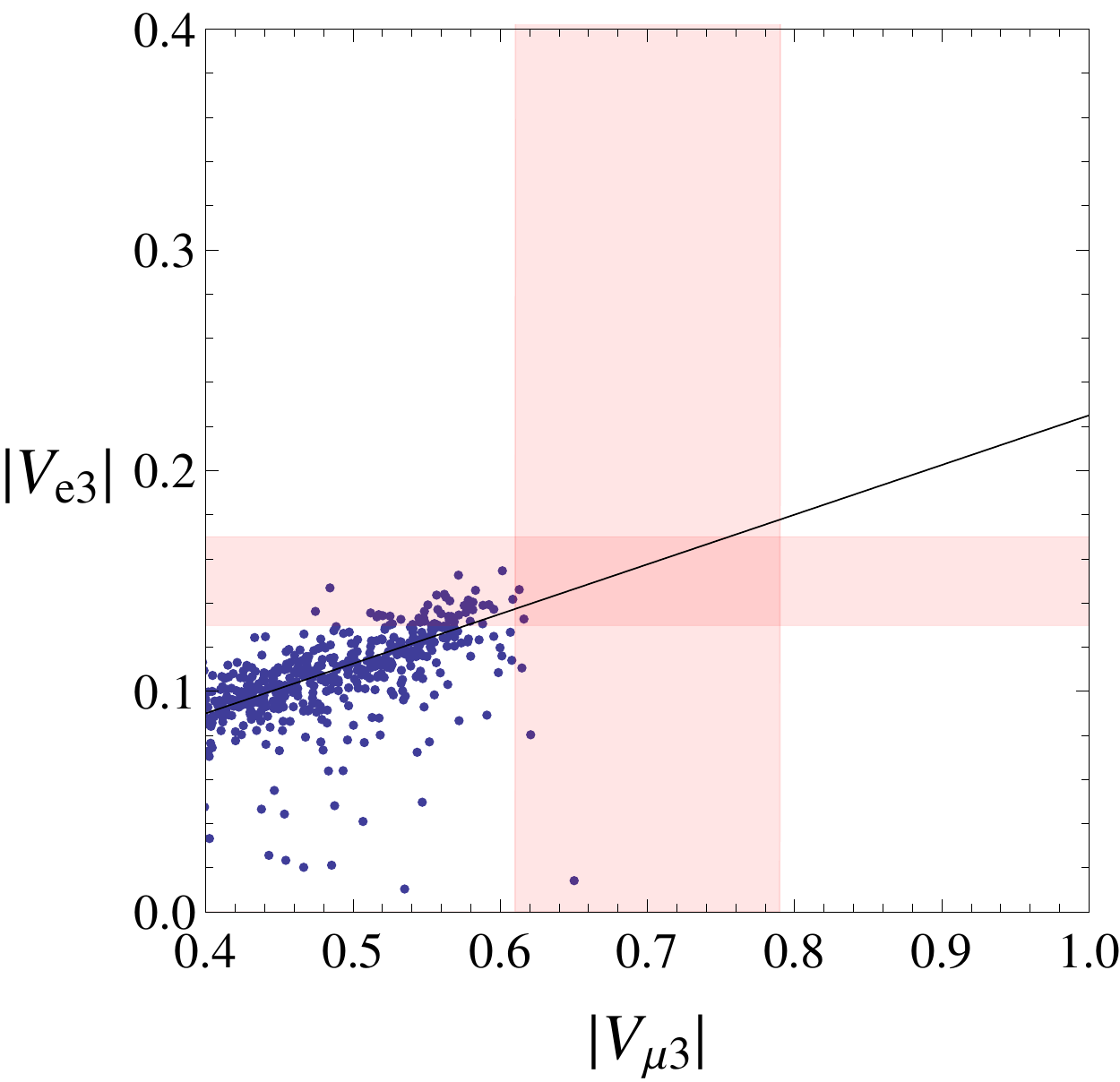} \\
               \includegraphics[height=2in]{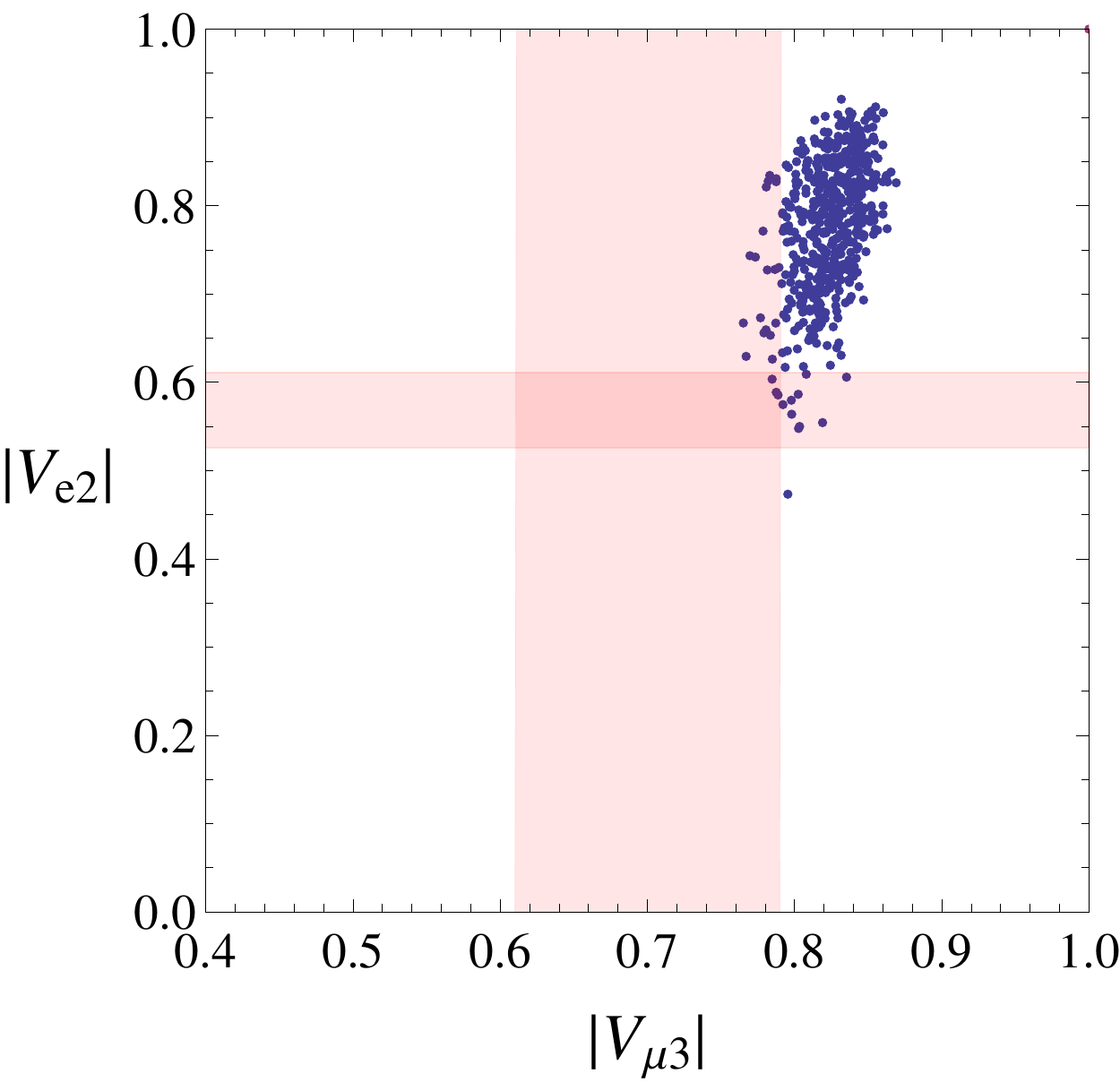}
               \includegraphics[height=2in]{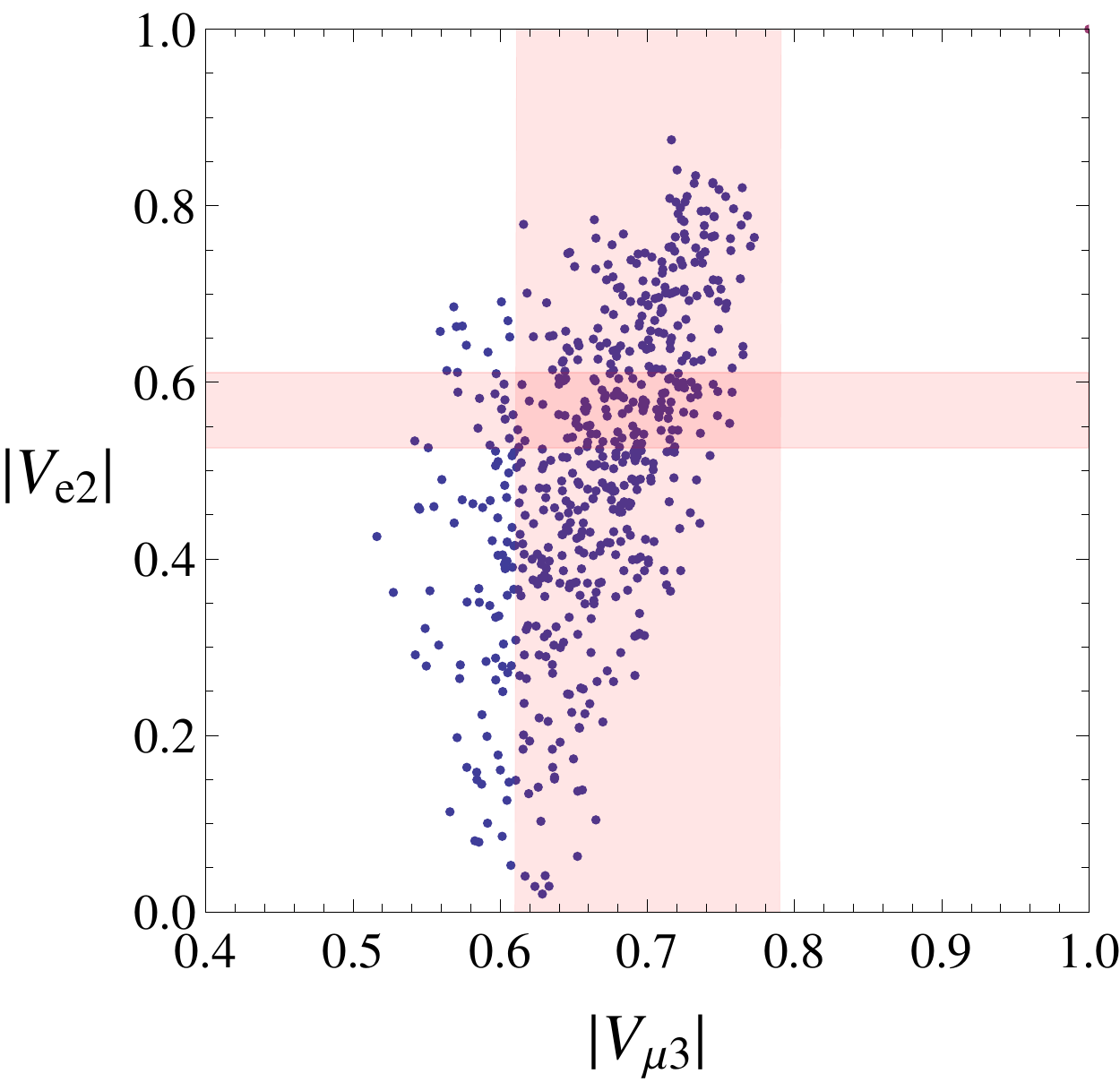} 
               \includegraphics[height=2in]{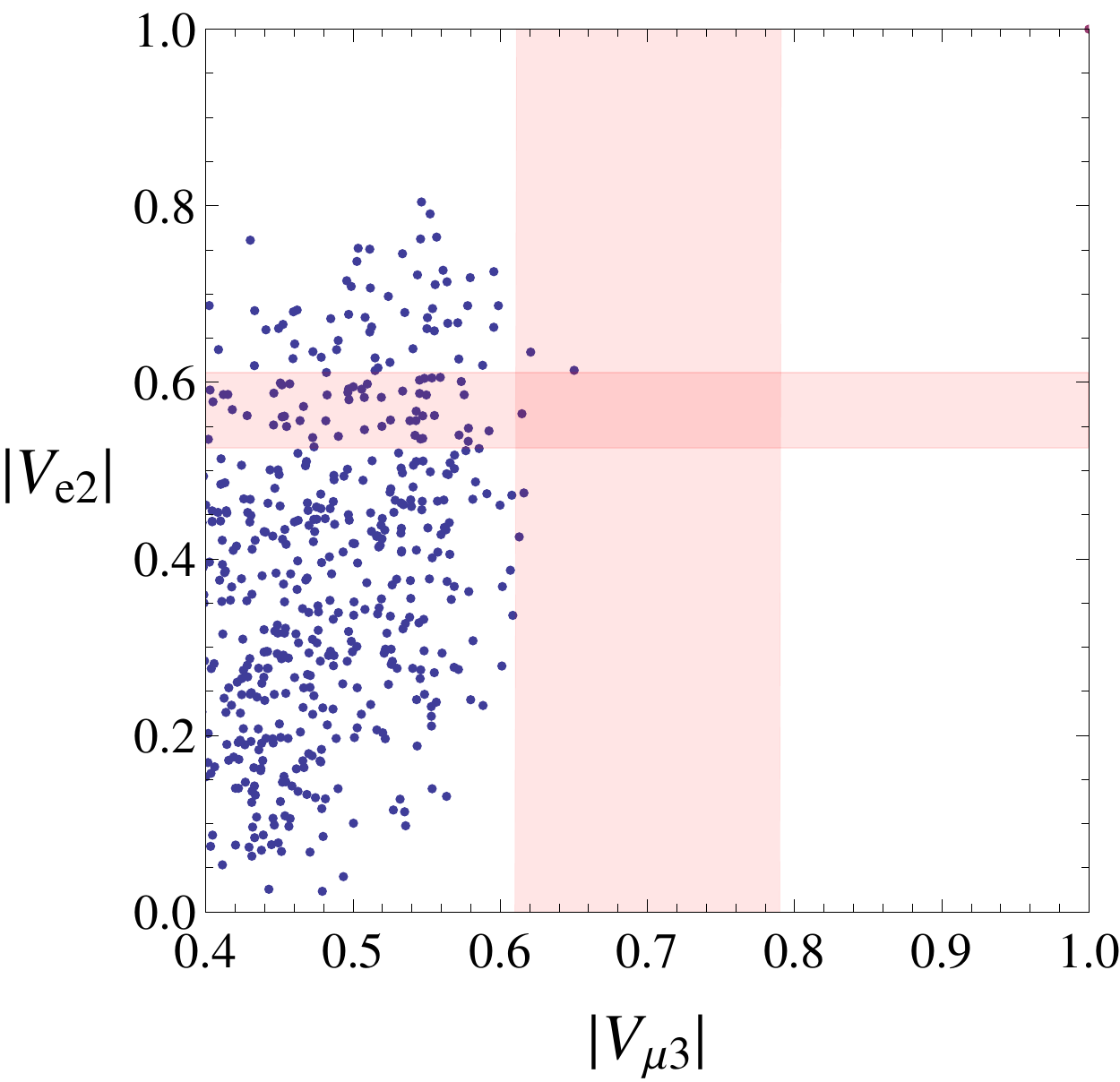}
 \end{center}
\vspace{-1cm}
                \caption{Scans of $|V_{e3}|$ versus
                  $|V_{\mu3}|$ (upper panels) and  $|V_{e2}|$ versus
                  $|V_{\mu3}|$ (lower panels) using the 2-3 symmetric Yukawa couplings of
                  Eq.~(\ref{23Y}), with $|d|=1/7$ and random $b, c,
                  e, g$, and their phases. Small random general
                  perturbations of order $5\%$ to these terms are
                  also included. The right-hand neutrino $c$-parameters
                  are also taken randomly, within three different
                  windows, {\it i.e.}, $c_{\nu}^i \in [-2.5,-2.3]$ (left
                  panels), $c_{\nu}^i \in [-2.2,-2.0]$ (center panels) and
$c_{\nu}^i \in [-2.0,-1.8]$ (right panels). The points show
                  agreement with the nontrivial correlation of
                  Eq.~(\ref{e3mu3})  (represented with the diagonal line in the upper graphs).
                }  
 \label{fig:ve3vmu3}
\end{figure}
Combining  
these equations we obtain the
relations linking PMNS and CKM mixing matrix elements:
\bea
V_{e3} &\simeq& - { V_{\mu3}} {V_{us}} - \frac{ V_{ub}}{\sqrt{2}} + {\cal
  O }(\epsilon), \label{e3mu3}\\
 V_{e2} &\simeq& \frac{1}{\sqrt{2}} \left (1-\frac{V_{ub}}{\sqrt{2}} \right) -V_{us} \left (1+\frac{V_{\mu 3}}{\sqrt{2}} \right) + {\cal
  O }(\epsilon) \label{e2mu3}.
\eea
In order to check the robustness of these relations we perform a
random scan of the free parameters in the Yukawa couplings,  
fixing only the
absolute value of $|d|=1/7$ and allowing its phase and the rest of
complex parameters of Eq.~(\ref{23Y}) $a,c,e,g$,  to be random, with absolute values
of ${\cal O}(1)$. The quark and charged lepton $c$-parameters are
fixed 
to obtain good first order values of the CKM angles and of masses, whereas in the neutrino
sector we take random values of $c_{\nu}^1$, $c_{\nu}^2$ and
$c_{\nu}^3$, with the only constraint that the values are in the
``neutrino plateau'' (see Fig.~\ref{fig:c})  and that their values remain relatively
degenerate. In particular we  
consider three windows, where we randomly
scan, {\it i.e.},  $c_{\nu}^i \in [-2.5,-2.3]$, $c_{\nu}^i \in [-2.2,-2.0]$ and
$c_{\nu}^i \in [-2.0,-1.8]$. During the scan over random parameters, we keep only points that
produce numerically correct CKM angles $|V_{us}|\simeq 0.22$,
$|V_{cb}|\simeq 0.041$, and $|V_{ub}|\simeq 0.0035$, as well as correct
neutrino mass differences. In
Fig.~\ref{fig:ve3vmu3}, we test the correlation between $|V_{e3}|$ and
$|V_{\mu3}|$ and find  it in very good agreement with
Eq.~(\ref{e3mu3}). The bands in the graph represent the $3\sigma$
uncertainty around the central values obtained from experimental
global fits
\cite{Gonzalez-Garcia:2014bfa,Maltoni:2004ei}. 
 It is quite remarkable that the theoretical
correlation curve as well as most of the points generated for the
intermediate window lie within these
bounds. Higher values of $c_{\nu}^i$ and lower values of $c_{\nu}^i$ produce
points with too high or too small $|V_{\mu3}|$.\footnote{Breaking further the
degeneracy in $c_{\nu}^i$ will increase the range of possible results.} Note that the lower values of
$c_{\nu}^i$ lie at the end of the plateau and the beginning of the exponential
sensitivity, and so these points start to show greater deviations from
the expected correlation. For higher $c_{\nu}^i$ values however, the correlation is
expected to be quite robust, although the experimental values might
not agree with the obtained results. 
\begin{figure}[!tb] 
\begin{center}
               \includegraphics[height=2.2in]{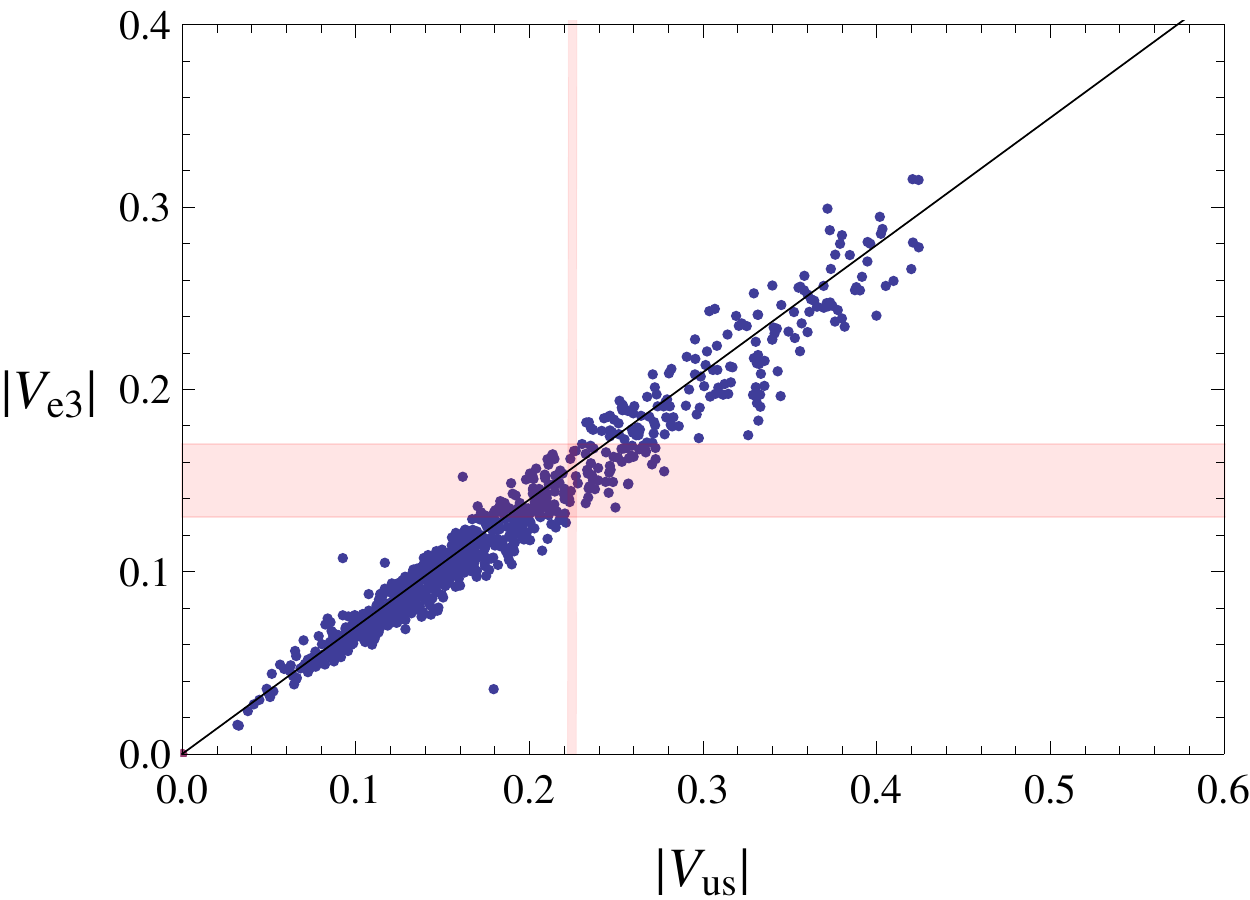}
\end{center}
\vspace{-1cm}
                \caption[Scan of the $V_{us}$ versus
                  $V_{e3}$]{Scan of the $V_{us}$ versus
                  $V_{e3}$ with random perturbations around 
                   2-3 symmetric Yukawa couplings as in the previous
                   figures, with  $c_{\nu}^i \in [-2.2,-2.0]$, and  
                  allowing for unphysical CKM entries (although still hierarchical), in
                  order to test the quark lepton correlation expressed
                  in Eq.~(\ref{e3us}) and shown in the graph with the
                  diagonal line.} 
                  \label{fig:vusve3}
\end{figure}
In order to further check this scenario, we  
perform the same scan
as before but  
including all the points generated, without checking for correct CKM values (although in
general they are quite CKM-like). This means that sometimes, the value of the 
Cabibbo angle is 
larger or smaller than expected due to
accidental alignments or suppressions coming from the Yukawa
couplings (taken randomly). Nevertheless the expectation is that the
correlation between $V_{e3}$ and $V_{us}$ from Eq.~(\ref{e3us}) should
survive, and indeed we see in Fig.~\ref{fig:vusve3} that this is the case. 
 The exact values of the parameters used in the scan are obtained by
adding small deviations to the zero-order values. In the
case of the $c$-parameters we have thus
$c_{f}^i=0.6\ \pm\ {\cal{O}}(0.1)$ for quarks and leptons, and
$c_{\nu}^i=2\ \pm\ {\cal{O}}(0.1)$ for right-handed neutrinos.
The Yukawa couplings are those given in Eq.~(\ref{23Y}), with the addition of
small random perturbations. 


\section{Flavor democracy}
\label{sec:03}
In the second symmetry implementation, in this section we assume a democratic structure \cite{Harari:1978yi,Pakvasa:1977in,Pakvasa:1978tx,
Derman:1979nf,Yamanaka:1981pa} for all  flavor parameters, meaning
that in our case the 5D Yukawa couplings, $Y_F^0$ are invariant under 
$S_3 \times S_3$  symmetry, while the 5D fermion bulk mass matrices, ${\mathbf c}^0_f$
are invariant under $S_3$ permutations. Explicitly, the democratic 5D Yukawa
couplings and 5D fermion bulk mass matrices are given by
\bea
Y^{Dem}_{F} \propto
\left(\begin{array}{ccc}
1 & 1 & 1 \\
1 & 1 &  1\\
1 & 1 & 1\end{array}\right) \ \  {\rm and}\ \
{\bf c}^{Dem}_{f}=\!\! \left(\begin{array}{ccc}
 a_f & b_f & b_f \\
b_f &  a_f &  b_f\\
b_f & b_f &  a_f\end{array}\right)\,. \ \ \  \label{democracy}
\eea
Overall, there are four Yukawa matrices, $Y^{Dem}_u$, $Y^{Dem}_d$,
$Y^{Dem}_e$ and $Y^{Dem}_\nu$, corresponding to the up-quark sector,
the down-quark sector, the charged leptons and the neutrinos. 
There are six fermion bulk $c$-matrices, namely ${\bf c}^{Dem}_{q}$,
${\bf c}^{Dem}_{u}$,  ${\bf c}^{Dem}_{d}$,   ${\bf c}^{Dem}_{l}$, 
${\bf c}^{Dem}_{e}$ and ${\bf c}^{Dem}_{\nu}$.  

Both matrices can be simultaneously diagonalized  
by the same unitary transformation
resulting in two zero eigenvalues for the Yukawa matrices, and two
degenerate eigenvalues for the bulk mass matrices, ${\mathbf 
  c}_f^{Dem}$. The 5D Yukawa and bulk mass matrices thus become, in their diagonal basis
\bea
Y^{0}_{F}= y^0_{{}_F} \left(\begin{array}{ccc}
0 & 0 & 0 \\
0 & 0 &  0\\
0 & 0 & 1\end{array}\right) \  {\rm and}\ \
{\bf c}^{0}_{f}=\!\! \left(\begin{array}{ccc}
 c^{1^0}_{f} & 0 & 0 \\
0 & c^{1^0}_{f}  &  0\\
0 & 0 & c^{3^0}_{f} \end{array}\right),\ \ \ \ \ \ \label{Ycdem}
\eea
where $y_F^0$ are complex Yukawa couplings and the index $F$ runs over $u$, $d$, $e$, and $\nu$.
The elements $c^{i\,0}_{f}$ are real and the index $f_i$ runs over doublets $q_i$, $l_i$ as well as singlets
$u_i$, $d_i$, $\nu_i$ and $e_i$, with $i$ the flavor index. Note that in this flavor symmetric
limit, all fermions except the $t$ quark, $b$ quark, $\tau$ lepton
and $\nu_{\tau}$ neutrino, are massless.
The 5D flavor structure of Eq. (\ref{Ycdem}) yields the 0-th order CKM
and PMNS matrices  
for this scenario
\bea\label{zerothPMNS}
V^0_{i}\!=
\left(\begin{array}{ccc}
\vphantom{\frac{9}{\sqrt{3}}}    \cos{\theta^0_{i}}   &
\sin{\theta^0_{i}}  &   \  0\  \\
\vphantom{\frac{\sqrt{9}}{\sqrt{3}}}    -\sin{\theta^0_{i}}  &  \cos{\theta^0_{i}}  &\   0\  \\
\vphantom{\frac{9}{\sqrt{3}}}         0       &     0        &   \ 1\
\end{array}\right)\, , \ \
\label{zeroorder}
\eea
where $i=$CKM, PMNS.  The angle $\theta_i^0$, depends on the detailed structure of the
symmetry breaking terms, $\delta Y_F$ and ${\delta \mathbf c}_f$
in Eqs.~(\ref{cexpansion}) and (\ref{Yexpansion}), and is not fixed by
the underlying $S_3\times S_3$ symmetry. 
Addition of generic small perturbations, as in Eqs.~(\ref{Yexpansion}) and (\ref{cexpansion}), 
 breaks the flavor symmetry and lifts the degeneracies to produce
SM-like masses and mixing angles. In the neutrino sector, the two level 
degeneracy is lifted by a small amount $(\delta Y_F)_{ij}$.
 This suggests a {\it normal} hierarchy
ordering with one heavier eigenstate, and with two lighter ones having similar
masses. Using Eq. (\ref{Mf}) and taking  
the generic size of 
the perturbations as $(\delta Y_F)_{ij}\simeq \delta Y^\nu$
for simplicity, yields the following relations for the neutrino masses
\bea\label{neutrinomass}
m_1& \sim  \delta Y^\nu\ v\ e^{-ky_1(a-1)} \ ,\qquad m_2
\sim  \delta Y^\nu\ v\ e^{-ky_1(a-1)} \ , \qquad m_3& \sim (1+\delta Y^{\nu})v \ e^{-ky_1(a-1)}.
\eea
Neutrino mass data requires  $v\ e^{-ky_1(a-1)} \simeq 0.3$ eV and 
   
explaining the hierarchy problem requires $ky_1\simeq 35$, which means
that the value of the Higgs localization parameter should be about $a
\simeq 1.8$. As explained in   
a previous section, this value of $a$
requires some fine-tuning of parameters in the RS 5D Higgs potential. 
  With modified $AdS_5$ metrics it is possible to remain
in a non-tuned region, and in particular we find that the best region
is obtained for $\nu\sim 0.2$ and $kL_1\sim 0.3$, where the parameter $a$ can
have values  
 as large as $4.5$.\footnote{This is due to the fact that in the
   modified $AdS_5$ case, Eq. (\ref{neutrinomass}) will be 
  slightly modified and higher values of the $a$-parameter become acceptable.}
 In order to obtain the observed neutrino mass hierarchy ratio $r$,
defined as $r=(|m_2|^2-|m_1|^2)/(|m_3|^2-|m_1|^2)\simeq 0.03$,  the size
of the Yukawa coupling perturbations $\delta Y$ must be fixed to $\delta Y^\nu
\lsim \sqrt{r}\simeq 0.17$.  There are no restrictions on the
values of bulk mass parameters $c_\psi^i$-s (as long as they are within the bounds $a<c_\nu+c_l$).

Consider the elements
$V_{e2}$, $V_{e3}$ and $V_{\mu3}$ of the PMNS matrix. As  mentioned
before, due to the plateau in the neutrino sector, for a small  $a$
parameter, the 5D Yukawa matrix structure is more or less preserved
and therefore these elements should be close to the ones shown in Eq.~(\ref{zerothPMNS}) plus some 
perturbations. 
This matrix predicts very small values for both
$|V_{e3}|$ and $|V_{\mu3}|$ and so the perturbations should lift them
(especially $|V_{\mu3}|$).
 The value of $V_{e2}$ on the other hand is highly sensitive to the
structure of the neutrino Yukawa flavor violating matrix  $\delta
Y^\nu_{ij}$ and can be large or small.  
\begin{figure}[!tb]
\bce
             \hspace{-1cm}   
             \includegraphics[height=6cm,width=5.5cm]{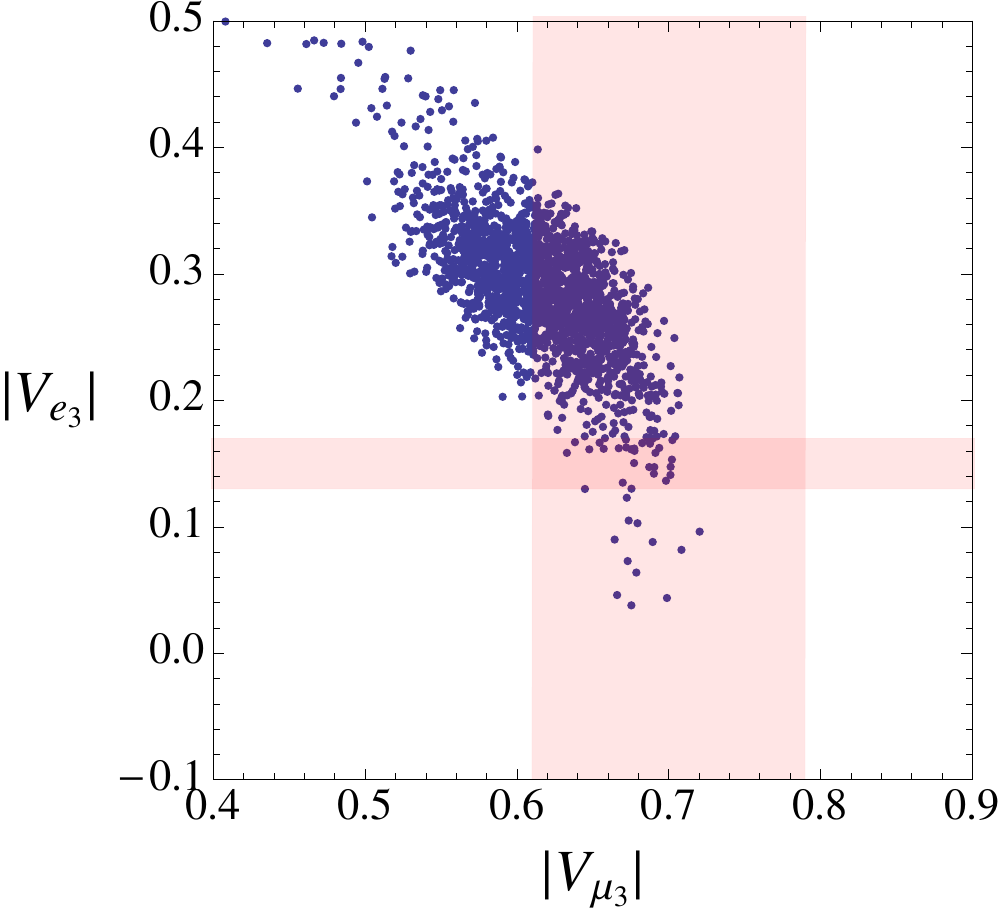}
                \includegraphics[height=6cm,width=5.5cm]{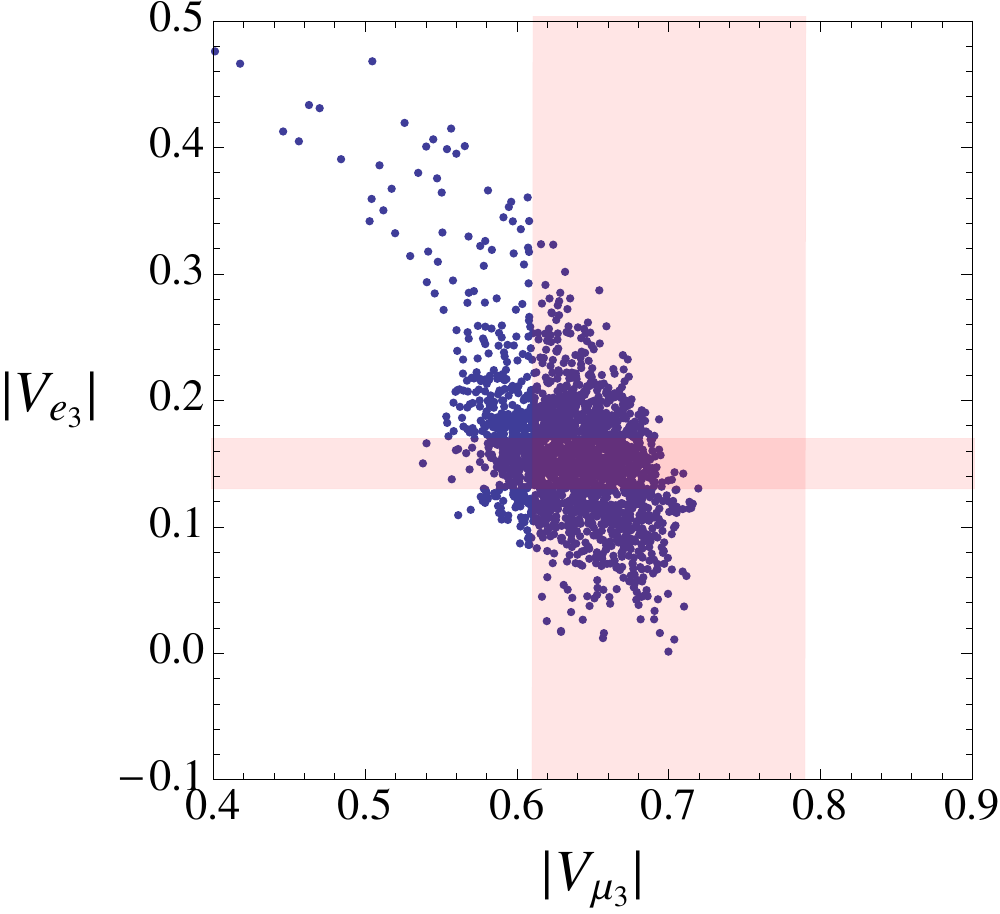}
                \includegraphics[height=6cm,width=5.5cm]{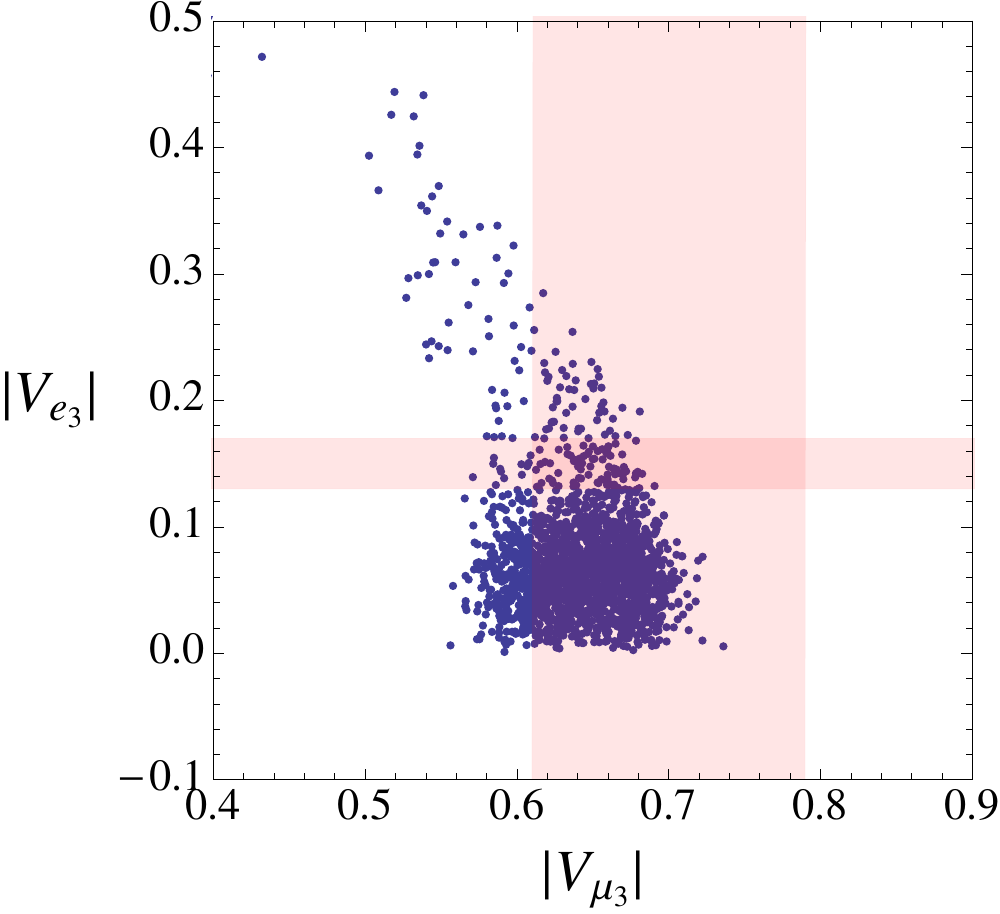}
\ece                
\vspace{-.4cm}
                \caption[Scatter plot of $V_{e3}$ versus $V_{\mu
                    3}$]{Scatter plots of $V_{e3}$ versus $V_{\mu 3}$
                  with random values of $\delta Y_{\nu}$, $\delta
                  Y_{e}$, $c_l$ and $c_{\nu}$. We have fixed $c_l^3=0.41$
                  and $(\delta Y^\nu)_{23}=0.13$ and then taken $(\delta
                  Y^\nu)_{13}=0.002$ (left panel) $(\delta
                  Y^\nu)_{13}=0.008$ (central panel) and  $(\delta
                  Y^\nu)_{13}=0.032$ (right panel). The concentration of points
                  in a precise region shows that the mixing angles
                  $V_{e3}$ and $V_{\mu 3}$ are sensitive to only these three
                  parameters. The obtained values of $V_{\mu 3}$
                  lie on the lower side of the experimental window,
                  whereas $V_{e3}$ depends on the ratio of Yukawa
                  couplings $(\delta Y^\nu)_{13}/ (\delta
                  Y^\nu)_{23}$, so that its smallness is due to a slight
                  hierarchy in these. }
 \label{fig:V3}
\vspace{.1cm}
\end{figure}
It turns out that when the c-parameters are such that,
$a<c_{l}^3+c_{\nu}^j$ and $c_{l}^3<1/2$ (a region not generic in usual
warped extra dimension scenarios) the lifting of the zero-order PMNS
can be successful and we have
\bea\label{PMNelem}
&&V_{e2}\sim  \sin{\theta}^0_{\nu} \ , \\
 && V_{e3}\propto \delta Y^\nu_{13}\ f(c^3_{l})\ ,\\ 
&&V_{\mu3}\propto \delta Y^\nu_{23}\ f(c^3_{l})\ .\ \ \
\eea
Note in particular that we expect that $V_{e3}/V_{\mu3} \propto
\delta Y^\nu_{13} / \delta Y^\nu_{23}$.
%
%
%
%
%
In Fig.~\ref{fig:V3} we present a scan of the model parameters to
verify the validity of Eq. (\ref{PMNelem}). In this scan we randomly
perturbed the values of the 5D parameters $Y_{\nu}$, $Y_{e}$, $c_l$ and 
$c_{\nu}$, but kept the values of the three relevant
parameters  $c^3_{l}$,  $\delta Y^\nu_{13}$ and $\delta Y^\nu_{23}$ fixed. We see from the figure that
the values obtained for the matrix elements $V_{\mu3}$ and $V_{e3}$ can
be made to lie within the experimental bounds by fixing only these three
parameters with all other terms randomly perturbed.
In particular, the formulas show the sensitivity of these two PMNS mixing angles to
the flavor structure of the neutrino Yukawa matrix $\delta Y^\nu$, but 
not to the charged lepton Yukawa matrix $\delta Y^l$ or to the bulk
masses $\delta c_\psi^i$, except for $\delta c^3_l$. Knowing that experimentally  $V_{\mu3}^{\rm exp}\simeq
0.65$ and $V_{e3}^{\rm exp}\simeq 0.15$, for the numerical evaluations
we took the bulk mass parameter of the third family lepton doublet
$c^3_{l}<1/2$  to obtain larger mixing angles for small $\delta
Y\simeq 0.1$, the same as
in the quark sector, where  $c^3_{q} \lesssim 1/2$ is needed  to
obtain a large top quark mass, and thus it could be a hinting of a possible additional
family symmetry among the $SU(2)$ doublets of the third family. The
particular values used in the scan are $c_l^3=0.41$, 
$\delta Y^\nu_{13}=0.008$ and $\delta Y^\nu_{23}=0.13$, which
produce PMNS angles consistent with experiment but with $|V_{\mu3}|$
on the lower experimental side. When the experimental uncertainty
decreases and if the central value of  $|V_{\mu3}^{\rm exp}|$ ends up
in the higher end of the current allowed region, this scenario will be
under great pressure. We also show how indeed the ratio
$V_{e3}/V_{\mu3}$ depends on  the ratio $\delta Y^\nu_{13} / \delta
Y^\nu_{23}$ by increasing and decreasing the value of  $\delta Y^\nu_{13}$ resulting in
larger or smaller values of $|V_{e3}|$.

In the charged lepton sector and the up- and down-quark sectors, massless
 states are also lifted by the flavor symmetry breaking, and these masses emerge directly
 proportional to the generic size of the perturbations in the Yukawa
 matrix, $\delta Y$, as described in \cite{Frank:2014aca}. 

Finally, in Table \ref{table1} we present a set of $S_3$ symmetric
0-th order bulk c-parameters (many other points with the same
symmetry in the parameter space are possible), which, with a small perturbation of
order  $\delta c\simeq\delta Y \simeq10\%$, can lead  to the
SM in the modified $AdS_5$ scenario\footnote{A similar point for the
  pure $AdS_5$ was presented in \cite{Frank:2014aca}.}. 
For this specific point we have taken the value of the warp exponent at the IR brane,
$A(y_1)$ to be exactly 35. With this assignment, there is only one
more free parameter in the model left to  completely fix the
metric. This parameter can be either $\nu$, or the position of the
singularity, $y_s$. Therefore one has the freedom to choose 
the amount of departure from the pure $AdS_5$. For the point presented in Table \ref{table1}, we took
$ \nu \simeq 0.32$ and the value of the localization of the Higgs field operator
$a$ is $4.46$ as in the previous example.
 With this starting point, the SM is then easily reproduced by breaking the symmetry 
 through adding 
perturbations $\delta c^i$-s and $\delta Y_{ij}$-s to lift the degeneracies of the symmetric scenario.
We require  $\delta Y_{ij}\lesssim 0.1$ and $\delta c_\psi^i \lesssim 0.1$
parameters to produce the SM masses, angles and phases,  by systematically using the
approximate formulas in this section up to any order of precision consistent with the SM. In general the
sensitivity to the precise values of the 5D Yukawa couplings, $\delta  Y_{ij }$,
is minimal and randomly chosen matrices with $\delta  Y_{ij}\lesssim 0.1$  produce
all the required SM features.

%
%
\begin{table}[ht]
\begin{center}
\begin{tabular}{c|c c c c c c}
$f $& $q $&$ l$ & $u$ & $d $& $e $ & $\nu$ \\
\hline\hline
$c^{1^0}_{f}                          $ & 0.55 & 0.55 & 0.55 & 0.65 &  0.65 & 5.00 \\
$c^{3^0}_{f}                          $ & 0.45 & 0.45 & 0.45 & 0.65 & 0.65 & 3.00 
\end{tabular}
\end{center}
\vspace{-.4cm}
\caption[A point in the 0-th order 5D fermion $c$-parameter space for
  $\mu$-$\tau$ symmetry symmetry]{A parameter point  for democratic flavor symmetry in
  the localization parameters  
  space (out of many possible
  points) in the 0-th order 5D fermion $c$-parameter space, consistent
  with all the experimental and model constraints. For this point, we
  have set all the 0-th order Yukawa coefficients to be universal,
  $y^0_u=y^0_d=y^0_\nu=y^0_e=4.4$,  and the Higgs localization
  parameter to $a=2.1$. The modified $AdS_5$ metric parameters are
  $\nu=1.1$, $y_1=2.8\times 10^{-17}$, and $A(y_1)=35$.} 
\label{table1}
\end{table}


\section{Conclusion}
\label{sec:summary}
In this work, we have provided a warped extra dimensional framework in which all of the fermions, including neutrinos, are
treated on equal footing, and where the SM fermion
flavor structure can still emerge naturally out of a slightly broken
universal flavor symmetry. Essential for this scenario is that all the matter fields must be in the bulk
and, in particular the Higgs field should be as delocalized as possible from
the IR boundary. We have explored a general warped scenario in which the
metric is modified from the usual $AdS_5$ background. This setup has
the advantage of allowing lower KK masses ($\sim$ 1-2 TeV), while still
safe from precision electroweak tests and flavor bounds. For our study, 
the modified metric presents a further advantage, as the neutrino mass
generation in our framework is more natural. This is due
to a numerical accident by which the neutrino masses generated in the $AdS_5$
background are too small (at most $\sim 10^{-4}$ eV) in the parameter
region of interest (neutrino plateau), while in the modified metric neutrino
masses can be up to two orders of magnitude larger, in the same
qualitative region of parameter space.  A way out in $AdS_5$ would be to further delocalize the Higgs field, although requiring such a delocalized
Higgs to generically solve the hierarchy problem,  some degree of fine-tuning must be 
 introduced in the Higgs potential. These tensions disappear when we use a modified $AdS_5$ geometry, where 
 our flavor setup can be successfully implemented without further
 fine-tuning.

In the parameter region of interest (the neutrino
plateau) the effective 4D neutrino masses do not have exponential
dependence on the bulk mass parameters $c^i$, in contrast with quark and charged lepton masses.
 Once a universal flavor symmetry is slightly broken,  the SM flavor structure
emerges due to the inherent features of warped space models, {\it i.e.} the
wave function profiles of light quarks and charged leptons are
exponentially sensitive to the symmetry violating terms, 
resulting in masses and mixing controlled by  small flavor violating terms.
In the neutrino sector, in the plateau region with a highly
delocalized Higgs field, the wave functions are not exponentially
sensitive to Lagrangian parameters  and thus the original flavor symmetry is essentially preserved. 
 Overall results are similar in both RS and modified $AdS_5$ type of scenarios, indicating that
the precise nature of the flavor symmetry or the precise nature of the
metric solution is not crucial for the main property of the
scenarios.
  
For illustration, we chose two simple examples of different
symmetries which can provide an implementation for this mechanism. In the
first scenario, we study  quark lepton
complementarity associated with $\mu-\tau$ symmetry (or 2-3 symmetry), known to lead
to phenomenologically viable neutrino masses and mixings. Assuming equal Yukawa matrix couplings in the charged lepton and down-quark sector, and equal Yukawa matrix couplings in the neutrino and up quark sector, as well as identical localization for the doublet quark and lepton representations, mass differences emerge entirely from singlet representation localization. The predictions of this implementation are definite connections between PMNS and CKM matrix elements, as given by Eqs. (\ref{e3mu3}) and (\ref{e2mu3}). 
 
In the second example, flavor democracy, the 5D Yukawa
couplings for the fermions  are invariant under the $S_3 \times S_3$ symmetry,
while the 5D fermion bulk mass matrices are invariant under $S_3$
family permutation invariance.  
Small perturbations of  ${\cal O}(10\%)$ are enough to generate the full flavor structure of the SM in both
quark and lepton sectors, and the ratio of PMNS matrix elements can be simply expressed in terms of these perturbations. Explicit expressions appear in Eqs. (\ref{PMNelem}). Localization of the third lepton family follows localization of the third quark family, hinting at an additional quark-lepton symmetry. As distinctive predictions and correlations appear in each
implementation,  the model studied here would yield a very promising novel laboratory for
studying fermion flavor symmetries.

\acknowledgments
We thank NSERC for partial financial support under grant number SAP105354.


\section{Appendix: Explicit expressions for the field profiles}
\label{app4}
\appendix

Here we derive the explicit expressions for the fermion and Higgs profiles, as well as for the Yukawa couplings for the modified $AdS_5$ scenario. The fermion profiles are given by
\beq
q^{0,i}_L(y) = q^i_0 e^{(2-c_q^i) A(y)}\ , \qquad
u^{0,i}_R = u^i_0 e^{(c_u^i+2) A(y)}\ ,
\eeq
while the Higgs profile is
\beq
h(y) = h_0 e^{a k y}\ ,
\eeq
with
\beq
q^i_0 =  \sqrt{k}\epsilon^{{1\over 2} - c_q^i}f(c_q^i) \equiv  \sqrt{k} \mathfrak{f}(c_q^i)\ , \qquad
u^i_0 =  \sqrt{k}\epsilon^{{1\over 2} + c_u^i}f(-c_u^i) \equiv  \sqrt{k} \mathfrak{f}(-c_u^i)\ ,
\eeq
and  
\begin{eqnarray}
h_0&\equiv&\sqrt{k} e^{- (a-1) k y_s} \mathfrak{h}_0  \\
\!\!\mathfrak{h}_0 &=& \frac{1}{\sqrt{ky_s  (2(a-1) k y_s)^{-\frac{2}{\nu ^2}-1} \left(\Gamma
   \left(1+\frac{2}{\nu ^2},2 (a-1) k (y_s-y_1)\right)-\Gamma \left(1+\frac{2}{\nu ^2},2 (a-1) k y_s\right)\right)}},\nonumber
\end{eqnarray}
where
\bea\label{epsilon}
\epsilon = e^{-A(y)} = e^{-k y_1} \left(1-\frac{y_1}{y_s}\right)^{\frac{1}{\nu ^2}}\ .
\eea

The reduced profile functions $\mathfrak{f}(c)$ are defined as  ~ $\mathfrak{f}(c)\equiv \epsilon^{{1\over 2} -c} f(c)\ , \quad {\text {with}}$
\begin{eqnarray}
&& f(c)\equiv\\\nonumber
&& \!\!\frac{\epsilon^{c-{1\over 2}}}{\sqrt{ky_s e^{(1-2
   c) k y_s} ((1\!-\!2 c) k y_s)^{\frac{1\!-\!2 c}{\nu ^2}-1} \left(\Gamma \left(1-\frac{1\!-\!2 c}{\nu ^2},(1\!-\!2 c) k
   (y_s-y_1)\right)\!-\!\Gamma \left(1-\frac{1\!-\!2 c}{\nu ^2},(1\!-\!2 c) k y_s\right)\right)}}. \\\nonumber
\end{eqnarray}
Note that the Higgs and the fermion profiles are defined differently,  due to the specific
Higgs potential we have considered  \cite{Azatov:2009na}. We can now write down the most general form for the Yukawa couplings as
\begin{eqnarray}
y_{ij}^u = \tilde{Y}^u_{ij} h_0 f(c_q^i) f(-c_u^j)\ ,
\end{eqnarray}
where the $\tilde{Y}_{ij}$s are related to the 5D Yukawa couplings via the equation
\begin{eqnarray}
\tilde{Y}_{ij}^u &\equiv& Y_{ij}^u  \sqrt{k}\epsilon^{1-c_q+c_u} y_s e^{k y_s (a-c_q^i+c_u^j)} \left [ k y_s
   (a-c_q^i+c_u^j) \right ]^{\frac{c_u^j-c_q^i}{\nu ^2}-1}  \\\nonumber & &\left[  \Gamma
   \left ( \frac{c_q^i-c_u^j}{\nu ^2}+1,(a-c_q^i+c_u^j) k (y_s-y_1)\right )-\Gamma
   \left (\frac{c_q^i-c_u^j}{\nu ^2}+1,(a-c_q^i+c_u^j) k y_s\right ) \right ] \  .
   \end{eqnarray}
Before switching to the expressions in the RS metric, we use the asymptotic expansion of the incomplete Gamma function
\bea
\Gamma\!\left(a,z\right)\sim z^{a-1}e^{-z}\left(1+{a-1\over z}+{(a-1)(a-2)\over z^2}+\mathcal{O}(z^{-3})\right)\, .
\eea
Then  we get the following
asymptotic behavior, up to the first order in $\mathcal{O}(z^{a-1})$,  with $\epsilon$ defined as in Eq. (\ref{epsilon}) %
\begin{eqnarray}\label{asymp1}
f(c) &\sim& \sqrt{ {1-2c \over 1-\epsilon^{1-2 c}} }\ , \\
h_0 &\sim& \sqrt{2  (1-a) \over 1- e^{2aky_1}\epsilon^{2}} \ , \\
\tilde{Y}_{ij}^{u} &\sim& \frac{e^{a ky_1}\epsilon^{c^i_q-c^j_u}-1}{a-c_q^i+c_u^j}Y_{ij}^{u}\ .
\end{eqnarray}
Keeping the $\mathcal{O}(z^{a-2})$ term not only gives a much better approximation for the top and neutrino plateaux, but
it also gives a more transparent expression for these functions:
\begin{eqnarray}
f(c) &\sim& \epsilon^{c-{1\over 2}} \sqrt{ {(1-2c)ky_s\nu^2 \over ky_s\nu^2(\epsilon^{2 c-1}-1) - \epsilon^{2 c -1} + \left({y_s\over y_s-y_1}\right)} }\ ,  \\
h_0&\sim& \epsilon^{-1}e^{-aky_1}\sqrt{ {2  (1-a)^2 \ ky_s\nu^2 \over ky_s\nu^2(1-a)(\epsilon^{-2}e^{-2aky_1}-1) - \epsilon^{-2}e^{-2aky_1} + \left({y_s\over y_s-y_1}\right)} }\ .
\end{eqnarray}
From these formulas, valid for the general modified $AdS_5$ metric,
one can see that, taking the limits $\nu\rightarrow \infty$ and
$y_s\rightarrow \infty$ we obtain the expressions for the profiles in
the RS metric: 
\begin{eqnarray}
f^{RS}(c) &=& \sqrt{ {1-2c \over 1-\epsilon^{1-2 c}} }\equiv \epsilon^{c-{1\over 2}}\mathfrak{f}^{RS}(c)\ , \qquad
h_0^{RS} = e^{(1-a)ky_1}\sqrt{2(1-a) \over\epsilon^{2(a-1)}-1}\ , \\
\tilde{Y}_{ij}^{RS, u} &\equiv& \frac{\epsilon^{-(a-c_q^i+c_u^j)}-1}{a-c_q^i+c_u^j}Y_{ij}^{RS, u} \ , \qquad
y_{ij}^{RS,u} = \tilde{Y}_{ij}^{RS,u} h_0^{RS} f^{RS}(c_q^i) f^{RS}(-c_u^j) \ .
\end{eqnarray}
where $\epsilon \equiv e^{-k y_1}$.
%

%
%
%
%

\end{document}